\documentclass[fleqn,usenatbib]{mnras}
\usepackage{color, colortbl}
\usepackage{multirow}

\usepackage{xcolor}
\definecolor{burntorange}{rgb}{0.8, 0.33, 0.0}

\usepackage{natbib}
\usepackage{amsmath}
\usepackage{amssymb}
\usepackage{bm} 
\hypersetup{linkcolor=red,citecolor=burntorange,filecolor=cyan,urlcolor=magenta}
\usepackage{longtable}
\usepackage[T1]{fontenc} 
\usepackage{verbatim} 
\usepackage{orcidlink}

\newcommand{\ang}{\mbox{\AA} }

\newcommand{\EBVratioavg}{$\langle E(B-V)_{\mathrm{star}}\rangle/\langle E(B-V)_{\mathrm{gas}}\rangle$}

\newcommand{\galex}{\textit{GALEX}}
\newcommand{\hst}{\textit{HST}}
\newcommand{\rst}{\textit{RST}}
\newcommand{\euclid}{\textit{Euclid}}
\newcommand{\jwst}{\textit{JWST}}
\newcommand{\spitzer}{\textit{Spitzer}}
\newcommand{\herschel}{\textit{Herschel}}
\newcommand{\halpha}{$\mathrm{H}\alpha$}
\newcommand{\hbeta}{$\mathrm{H}\beta$}
\newcommand{\BD}{$\mathrm{H}\alpha/\mathrm{H}\beta$}
\newcommand{\zspec}{$z_\mathrm{spec}$}
\newcommand{\magphys}{\texttt{MAGPHYS}}

\title[Average dust attenuation curve at $z\sim1.3$]{The average dust attenuation curve at $z\sim1.3$ based on \hst\ grism surveys}

\author[A. J. Battisti et al.]{
A. J. Battisti\orcidlink{0000-0003-4569-2285}$^{1,2}$\thanks{Email:andrew.battisti@anu.edu.au},
M. B. Bagley\orcidlink{0000-0002-9921-9218}$^{3}$,
I. Baronchelli\orcidlink{0000-0003-0556-2929}$^{4}$,
Y.-S. Dai\orcidlink{0000-0002-7928-416X}$^{5}$,
A. L. Henry\orcidlink{0000-0002-6586-4446}$^{6}$,
\newauthor
M. A. Malkan\orcidlink{0000-0001-6919-1237}$^{7}$,
A. Alavi\orcidlink{0000-0002-8630-6435}$^{8}$,
D. Calzetti\orcidlink{0000-0002-5189-8004}$^{9}$,
J. Colbert$^{8}$,
P. J. McCarthy$^{10}$,
V. Mehta\orcidlink{0000-0001-7166-6035}$^{11}$,
\newauthor
M. Rafelski\orcidlink{0000-0002-9946-4731}$^{6,12}$,
C. Scarlata\orcidlink{0000-0002-9136-8876}$^{11}$,
I. Shivaei\orcidlink{0000-0003-4702-7561}$^{13}$,
E. Wisnioski\orcidlink{0000-0003-1657-7878}$^{1,2}$
\\
$^{1}$Research School of Astronomy and Astrophysics, Australian National University, Cotter Road, Weston Creek, ACT 2611, Australia\\
$^{2}$ARC Centre of Excellence for All Sky Astrophysics in 3 Dimensions (ASTRO 3D), Australia\\
$^{3}$Department of Astronomy, The University of Texas at Austin, Austin, TX 78712, USA\\
$^{4}$Dipartimento di Fisica e Astronomia, Universit\`a di Padova, Vicolo dell’Osservatorio, 3, 35122 Padova, Italy\\
$^{5}$Chinese Academy of Sciences South America Center for Astronomy (CASSACA), National Astronomical Observatories of China (NAOC), 20A Datun Road, Beijing, 100012, China\\
$^{6}$Space Telescope Science Institute, 3700 San Martin Dr., Baltimore, MD 21218, USA\\
$^{7}$Department of Physics and Astronomy University of California, Los Angeles Los Angeles, CA 90095-1547, USA\\
$^{8}$IPAC, California Institute of Technology, 1200 E. California Boulevard, Pasadena, CA 91125, USA\\
$^{9}$Department of Astronomy, University of Massachusetts-Amherst, Amherst, MA 01003, USA\\
$^{10}$National Optical-Infrared Astronomy Research Laboratory, Tucson, AZ 85719, USA\\
$^{11}$Minnesota Institute for Astrophysics, University of Minnesota, Minneapolis, MN 55455, USA\\
$^{12}$Department of Physics and Astronomy, Johns Hopkins University, Baltimore, MD 21218, USA\\
$^{13}$Steward Observatory, University of Arizona, 933 N Cherry Ave, Tucson, AZ 85721, USA\\
}

\date{Accepted XXX. Received YYY; in original form ZZZ}

\pubyear{2022}

\begin{document}
\label{firstpage}
\pagerange{\pageref{firstpage}--\pageref{lastpage}}
\maketitle

\begin{abstract}
We present the first characterisation of the average dust attenuation curve at $z\sim1.3$ by combining rest-frame ultraviolet through near-IR photometry with Balmer decrement (\halpha /\hbeta) constraints for $\sim$900 galaxies with $8\lesssim\log (M_\star /M_\odot)<10.2$ at $0.75<z<1.5$ in the \hst\ WFC3 IR Spectroscopic Parallel (WISP) and 3D-HST grism surveys. Using galaxies in SDSS, we establish that the (\halpha +[NII])/[OIII] line ratio and stellar mass are good proxies for the Balmer decrement in low-spectral resolution grism data when only upper-limits on \hbeta\ are available and/or \halpha\ is blended with [NII]. The slope of the $z\sim1.3$ attenuation curve ($A(0.15\micron)/A(V)=3.15$) and its normalization ($R_V=3.26$) lie in-between the values found for $z=0$ and $z\sim2$ dust attenuation curves derived with similar methods. These provide supporting evidence that the average dust attenuation curve of star forming galaxies evolves continuously with redshift. The $z\sim1.3$ curve has a mild 2175\AA\ feature (bump amplitude, $E_b=0.83$; $\sim$25\% that of the MW extinction curve), which is comparable to several other studies at $0<z\lesssim3$, and suggests that the average strength of this feature may not evolve significantly with redshift. The methods we develop to constrain dust attenuation from \hst\ grism data can be applied to future grism surveys with \jwst , \euclid , and \rst. These new facilities will detect millions of emission line galaxies and offer the opportunity to significantly improve our understanding of how and why dust attenuation curves evolve. \\
\end{abstract} 

\begin{keywords}
galaxies: evolution -- galaxies: ISM -- galaxies: star formation -- (ISM:) dust, extinction -- ISM: evolution
\end{keywords}

\section{Introduction}
In the near future, the synergy of large photometric surveys such as the Vera C. Rubin Observatory Legacy Survey of Space and Time (LSST), the \textit{Nancy Grace Roman Space Telescope} (\rst) , and \euclid\ will yield precise galaxy spectral energy distributions (SEDs) for billions of galaxies throughout cosmic time and move us into a regime of precision cosmology for both galaxy evolution and dark energy studies. However, dust within galaxies absorbs and scatters light (referred to as attenuation) in a complex manner that depends on the intrinsic properties of the dust (extinction\footnote{We define dust extinction as the absorption and scattering of light out of the line of sight by dust, which has no dependence on geometry. Dust attenuation is the combination of extinction, scattering of light into the line of sight by dust, and geometrical effects due to the star-dust geometry.} curve) and on the relative star/dust geometry \citep{calzetti01, salim&narayanan20}. A critical ingredient of future precision work is a description of dust attenuation curves\footnote{Two common methods to quantify dust curves are as total-to-selective extinction/attenuation, $k(\lambda)=A(\lambda)/(A(B)-A(V))$, or as a normalised curve, $k(\lambda)/k(V)=A(\lambda)/A(V)=\tau(\lambda)/\tau(V)$, where $A(\lambda)$ is the extinction/attenuation in units of magnitudes and $\tau(\lambda)$ is the optical depth. The $B$- and $V$-bands are usually assumed to be 4400~\AA\ and 5500~\AA, respectively.}, and how they evolve with time. This issue is especially pertinent at $1<z<3$ where galaxies are more heavily attenuated than in the local Universe \citep[e.g.,][]{magnelli13} and which corresponds to the peak in cosmic star formation \citep{madau&dickinson14}. 

Current approaches to correct for the effects of dust attenuation on galaxy SEDs are limited in scope and often rely on the average attenuation curve based on $\sim$40 local starburst galaxies \citep{calzetti00}.
However, using a sample of $\sim$200 galaxies at $z\sim2$, \citet{reddy15} derived an attenuation curve that has a notably lower total-to-selective attenuation in $V$-band, $R_V$ (same as $k(V)$), than that found for local galaxies ($R_V\sim4.0$ and $\sim2.5$ at $z=0$ and $z\sim2$, respectively) and also steeper UV curve slope $S=A(0.15)/A(V)$ ($S\sim2.5$ and $S\sim3.5$ at $z=0$ and $z\sim2$, respectively), suggesting significant variation as a function of redshift \citep[see][]{salim&narayanan20}. More recently, \citet{shivaei20a} extended the results of \citet{reddy15} to demonstrate that the UV slope of the dust curve and the strength of the 2175\ang feature depend on the gas-phase metallicity for $z\sim2$ galaxies. Studies of dust attenuation in local galaxies ($z\sim0$) are unable to account for all of these differences \citep[e.g.,][]{wild11, battisti17a, battisti17b, salim18}, which may stem from intrinsically different properties of the interstellar medium (ISM) of typical low-$z$ and high-$z$ galaxies (e.g., metallicity, dust composition). This highlights the need to characterise attenuation curves in a large sample of galaxies at $z\sim1$ to verify and improve our understanding of the evolution of dust attenuation curves from $z\sim2$ to $z\sim0$.

The main difficulty in improving our picture of dust attenuation at higher redshifts is due to the limited number of galaxies with spectroscopy ideal for its characterisation (e.g., Balmer decrements; \BD). The development of multiplexed near-IR spectrographs, such as Keck/MOSFIRE \citep{mcLean10, mcLean12}, have enabled progress on this front. In particular, the MOSFIRE Deep Evolution Field (MOSDEF) survey \citep{kriek15} obtained high-resolution spectroscopy of $\sim$1500 galaxies at $1.4\lesssim z\lesssim3.8$, of which $\sim$200 are individually detected in both \halpha\ and \hbeta\ to provide Balmer decrements. This survey was used by \citet{reddy15} and \citet{shivaei20a} to derive dust attenuation curves at $z\sim2$.

Another fruitful avenue comes from slitless grism surveys with the \textit{Hubble Space Telescope} ($HST$), such as the WFC3 Infrared Spectroscopic Parallel \citep[WISP, PI: M. Malkan][]{atek10} and 3D-HST \citep{brammer12} surveys, which provided low-resolution spectroscopy ($R\approx200$) for thousands of galaxies at $0.3\lesssim z\lesssim 2.3$. Two disadvantages to these surveys are that only a small fraction of the entire sample has detected \hbeta\ at $S/N>3$, and their low spectral resolution introduces line blending (e.g., \halpha +[NII]).  However, by using stacking analyses of large numbers of galaxies, together with calibrated corrections to account for blending, it is now possible to overcome these shortcomings and characterise dust attenuation curves at $z\sim1$ using grism data. 

In this paper, we construct a sample of $\sim$1200 emission line galaxies (ELGs) at $0.75<z<1.5$ from the 3D-HST and WISP surveys to characterise the average attenuation curve for galaxies at $z\sim1.3$. These ELGs have both \halpha\ and \hbeta\ covered in the grism spectroscopy to measure Balmer decrements and SED coverage from rest-frame UV to near-IR. The methods and results presented in this work are intended to act as a benchmark for future studies with \jwst, \euclid, and \rst, which all have near-IR grism capability (and similar spectral resolution to \hst), and will dramatically increase the number of ELGs that are detected at higher redshifts \citep[e.g.,][]{bagley20}. These large statistical datasets will provide the opportunity to explore the nature of dust attenuation curve variation and evolution and establish if such changes can be attributed to galaxy properties.

Throughout this work we adopt a $\Lambda$-CDM cosmological model, with $H_0=70$~km/s/Mpc, $\Omega_M=0.3$, and $\Omega_{\Lambda}=0.7$.

\section{Data and Sample}\label{data_sample}
Our data consist of a combination of photometric and grism spectroscopic data from \hst\ and ancillary photometric surveys. The WISP and 3D-HST programs performed near-infrared slitless spectroscopic observations using one or both of the WFC3 IR grisms: G102 (0.85-1.10~\micron , $R\sim210$) and G141 (1.07-1.75~\micron , $R\sim130$), as well as imaging in the equivalent filter(s) ($F110W$, $F140W$, or $F160W$). In order to characterise the dust attenuation curve from these samples, it is necessary to combine this spectroscopic information with photometric data covering rest-frame UV through near-IR wavelengths. Here we summarise the data available for each survey. 

All photometric data are extinction corrected for foreground Milky Way dust using dust maps from \citet{schlafly&finkbeiner11} and assuming the MW extinction curve of \citet{fitzpatrick99}. 

\subsection{WISP Survey}\label{WISP_summary}
The WFC3 Infrared Spectroscopic Parallel  \citep[WISP, PI: M. Malkan;][]{atek10} survey is a multi-cycle \hst\ pure-parallel program that obtained WFC3 observations for 483 pointings in random extragalactic fields (i.e., location depended on the primary observing target). Due to the nature of parallel observations, the integration times for each field was set by the primary target observations. We refer the reader to \citet{atek10} for a complete description of the observing strategy. In brief, short opportunities (one to three continuous orbits) typically obtain G141 grism along with one imaging filter (F140W or F160W) and long opportunities (four or more continuous orbits) obtain G102+G141 grisms and F110W+F160W (or F140W) imaging. For the long opportunities, the integration times in the two grisms were set to achieve approximately uniform sensitivity for an emission line of a given flux across the full wavelength range. The median $5\sigma$ detection limit for emission lines fluxes (point source) in both grisms is $~5\times 10^{-17}$~erg~s$^{-1}$~cm$^{-2}$ \citep{atek10}. For $\sim150$ of the long opportunity fields, WFC3/UVIS imaging data was also obtained with some combination of F475X, F606W, F600LP, and/or F814W filters (only 2 of these at most for a single field). 
For this paper, the WISP parent sample includes emission-line measurements from 419 WISP fields ($\sim$1520~arcmin$^2$; Bagley et al. in prep.). For the deep fields, the G102+G141 data from WISP provides spectral coverage of the Balmer decrement from $0.75<z<1.5$.

A strength of the WISP survey is that the fields are independent and uncorrelated (i.e., unbiased). However, due to the parallel nature of WISP, these fields do not typically have deep ancillary photometric data available. The WISP team has carried out several supplementary observing programs to obtain additional photometry with a variety of facilities, with priority given to the deepest $\sim$200 fields. These include optical data ($ugri$) from Palomar/LFC (numerous PIs: I. Baronchelli, J. Colbert, S. Dai, M. Rafelski, B. Siana, C. Scarlata, H. Teplitz), Palomar/WaSP (PI: S. Dai), WIYN/MiniMosaic (PI: A. Henry), WIYN/ODI (PI: A. Battisti), and Magellan/Megacam (PIs: A. Battisti \& P. McCarthy), as well as near-IR (3.6\micron) from \spitzer /IRAC ch1 (PI: J. Colbert). However, we note that a significant fraction ($\sim$60\%) of the WISP fields have little or no ancillary photometry and therefore are not usable for this study to characterise dust attenuation curves. 
 
The details and public release the \hst\ data will be presented in Bagley et al. (in prep.). In brief, all \hst\ images are drizzled onto the same pixel scale, optimised for WFC3/UVIS (0.04"/pixel), and then convolved with a kernel to match the point spread-function (PSF) in the F160W filter. A segmentation map is generated from the F110W and F160W detections using \texttt{SExtractor} \citep{bertin&arnouts96}. The photometry is derived using \texttt{photutils} in Astropy to derive isophotal fluxes in all \hst\ bands \citep{astropy13, astropy18}, with local sky subtraction. \texttt{SExtractor} photometry is also performed on WFC3/IR (0.08"/pixel) images (prior to PSF-matching), in order obtain the total magnitudes (\texttt{MAG\_AUTO}). Aperture corrections on the isophotal data are derived from the difference between \texttt{MAG\_AUTO} and the isophotal magnitude in the reddest \hst\ filter (usually F160W). The median value of these aperture corrections (\texttt{MAG\_AUTO}-\texttt{MAG\_ISO}) for our WISP sample is $\Delta m=-0.11$~mag.

The details and public release of the WISP ground-based optical data will be presented in Battisti et al. (in prep.) and the \spitzer\ data will be presented in Baronchelli et al. (in prep.). For both of these (lower-resolution) datasets, self-consistent, total photometry is obtained using the \texttt{TPHOT} software \citep[v2, ][]{merlin16}. \texttt{TPHOT} performs PSF-matched, prior-based, multiwavelength deconfusion photometry to optimally extract photometric measurements or upper limits for lower resolution and/or shallower datasets. The reddest \hst\ filter (usually F160W) is adopted as the prior for \texttt{TPHOT} source extraction. Our final combined catalog contains PSF-matched photometry in all filters.
 
As part of the standard WISP emission line pipeline, all spectra were visually inspected and interactively fit by at least two people for reliability. The pipeline produced emission line catalog (Bagely et al. in prep.) provides us with a parent sample of 1294 galaxies with the necessary spectroscopic data and coverage, but only 314 galaxies that also satisfy our photometric selection criteria (Section~\ref{selection}). For our study, we re-examine the spectral fits for all candidate sources in the redshift window where both \halpha\ and \hbeta\ are covered to check the following: (1) reassess that the fits to emission lines are reasonable, (2) examine the continuum fit around the \hbeta\ region and refit this if necessary by modifying/adding to the spline nodes used for the continuum shape, (3) ensure that transition point between the G102 and G141 spectra, which slightly overlap, is suitable for a good fit of the continuum region on either side of \hbeta\ and [OIII]. We note that due to the sheer number of sources that were visually inspected as part of the WISP pipeline process, it was not a priority for the WISP assessors to examine the quality of the continuum fits around regions of undetected lines, warranting this re-examination. None of the 314 sources re-examined were rejected due to contamination or an ambiguous redshift solution.


\subsection{3D-HST Survey}\label{3DHST_summary}
The 3D-HST survey \citep[PI: P. van Dokkum]{brammer12} was a multi-cycle \hst\ Treasury program that obtained WFC3 observations of the CANDELS fields \citep[AEGIS, COSMOS, GOODS-N, GOODS-S, UDS]{grogin11, koekemoer11}. We note that most of GOODS-N was observed as part the previous AGHAST program \citep[PI: B. Weiner]{weiner09}. These data covered $\sim$150 \hst\ pointings (626~arcmin$^2$) to a uniform two-orbit depth with the G141 grism and F140W (direct imaging), achieving a $3\sigma$ limit for emission line fluxes (point source) of $2.1\times 10^{-17}$~erg~s$^{-1}$~cm$^{-2}$ \citep{momcheva16}. The reduced 3D-HST spectroscopic data for all galaxies detected in imaging is presented in \citet{momcheva16}. The 3D-HST pointings reside in legacy fields that also have extensive ancillary photometric data that provide rest-frame UV through near-IR coverage for all of the fields. The photometric catalog of this ancillary data, along with several derived redshifts and physical properties, is presented in \citet{skelton14} and is used for our analysis. When available, we also include IR photometry from \spitzer\ 24\micron\ and \herschel\ from \cite{whitaker14} and \cite{wuyts16}, respectively. In the context of this study, the single-grism aspect of 3D-HST limits our selection to galaxies at $1.3<z<1.5$ (i.e., the Balmer decrement window). 

To improve the reliability on the redshift and emission line fit for single-line emitters, we require that the 2$\sigma$ confidence interval of the photometric redshift in the 3D-HST public catalog \citep{skelton14} overlap with the spectroscopically derived redshift. This provides us with a parent sample of 1443 galaxies in 3D-HST with $S/N$(\halpha+[NII])>3 (in the 3D-HST emission line catalog) to be considered. The 3D-HST spectroscopy is reprocessed to be similar to the format utilised for the WISP emission pipeline (Bagley et al. in prep.) and follows the method presented in \citet{henry21}. As a result of this procedure, all 3D-HST spectra are visually inspected and interactively fit in the exact same manner as done for the WISP sample and provides self-consistency between the datasets. 

The public 3D-HST data products used an automated fitting procedure. During the interactive fitting, 182 out of 1443 candidates (13\%) were rejected and removed due to either being contamination or considered spurious line detections. We highlight that we took particular care to fit the continuum shape in the \hbeta\ region as this is very important for the reliability of the stacking procedure that is the backbone of our analysis. We obtain a sample of 864 galaxies that satisfy our selection criteria (Section~\ref{selection}), noting that most sources removed are due to having $S/N$(\halpha+[NII])<3 after being run through the WISP pipeline. For \halpha+[NII], we find that that the average $S/N$ values from the 3D-HST catalog are roughly 30\% larger than those inferred from the WISP pipeline. A small number were also slightly outside our desired redshift range (i.e., to require a slight buffer on both sides of Balmer lines for accurate continuum). 
 
\subsection{Sample Selection Criteria}\label{selection}
We select a robust sample of 1178 galaxies from WISP and 3D-HST by requiring that the following conditions are met:
\begin{enumerate}
\item[(1)] Grism spectral coverage of both \halpha +\hbeta 
\item[(2)] $S/N(\mathrm{H}\alpha\mathrm{+[NII]}) \ge 3$ and one additional line with $S/N \ge 2$
\item[(3)] Sufficient SED coverage from photometry
\subitem (a) $\lambda_\mathrm{rest}<0.30\micron$ in at least two filters
\subitem (b) $\lambda_\mathrm{rest}>1.0\micron$ in at least one filter
\subitem (c) At least five bands with $S/N>3$
\end{enumerate}
Condition (1) ensures that we can recover Balmer decrements from stacking. Condition (2) ensures an accurate \zspec , which is important for optimally aligning the spectra, as well as reducing false identifications. For example, \citet{baronchelli20} show that the default choice of assuming single-line emitters are \halpha\ in WISP is incorrect for $\sim$30\% of cases, where most of these are likely to be the [OIII] emission line. Condition (3) ensures that we have adequate SED coverage for characterising dust attenuation curves and to recover reliable UV-slopes ($\beta$) and stellar masses, $M_\star$. The five band requirement removes a handful of galaxies in the WISP sample that have only one band in the rest-frame optical. There are always more than five bands detected for galaxies in the 3D-HST sample. 

\begin{figure}
\begin{center}
\includegraphics[width=0.45\textwidth]{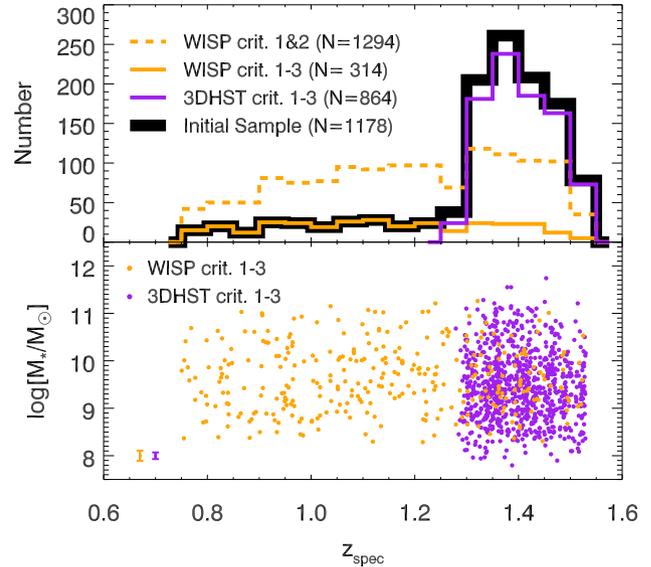}
\end{center}
\vspace*{-0.5cm}
\caption{The top panel shows the redshift distribution for galaxies in the WISP (orange) and 3D-HST (purple) samples that satisfy the criteria described in Section~\ref{selection}. The black solid lines shows our initial sample satisfying all criteria (1 - 3), which has a mean of $\langle z\rangle=1.33$. A significant fraction of the WISP fields do not currently have the necessary ancillary photometric data to be included in this study (orange dashed line). The bottom panel shows stellar masses, derived from \magphys\ (Section~\ref{method_SED_fitting}), as a function of redshift. The median errorbar on stellar mass for each survey is indicated (0.11~dex for WISP, 0.07~dex for 3D-HST). No significant selection effect on stellar mass is apparent with redshift or between the two surveys. 
 \label{fig:z_hist}}
\end{figure}

The redshift distribution of our sample satisfying all of these criteria is shown in Figure~\ref{fig:z_hist}. We note again that the WISP sample has a wider Balmer decrement window (G102+G141) than 3D-HST (G141), but that 3D-HST has deeper and more uniform ancillary photometric data relative to the WISP survey (see Section~\ref{WISP_summary} and \ref{3DHST_summary}). The conditions listed above restrict us to an initial sample size of 314 and 864 galaxies in WISP and 3D-HST, respectively. We note that if sufficient photometric coverage was available for all of the WISP fields, its number would increase to $\sim$1300 usable galaxies (i.e., photometry is the primary limiting factor to using a larger fraction of the WISP sample). We note that only $\sim$10.8\% of our initial sample has \hbeta\ detected at $S/N>3$.

Additional minor cuts on this sample are imposed after SED-fitting, and are based on the goodness of fit, the accuracy of stellar masses, the accuracy of UV slopes, and to exclude active galactic nuclei (AGN) candidates. These requirements are described in more detail in Section~\ref{method_SED_fitting} and \ref{method_agn_selection}. After these additional cuts, we are left with a final sample size of 1081 galaxies for analysis, where 272 are from WISP and 809 are from 3D-HST. 

\subsection{SDSS Comparison Sample ($z<0.2$)}\label{data_sdss}
In order to provide supporting evidence for the choices made in our stacking analysis, we compare to a sample of local ELGs from the Sloan Digital Sky Survey (SDSS) data release 7 \citep[DR7;][]{abazajian09}. We use the optical spectroscopic and physical property measurements for these galaxies from the Max Planck Institute for Astrophysics and Johns Hopkins University (MPA/JHU) group\footnote{\url{http://www.mpa-garching.mpg.de/SDSS/DR7/}}. The stellar masses are based on fits to the photometric data following the methodology of \citet{kauffmann03a} and \citet{salim07}. The SFRs are based on the method presented in \citet{brinchmann04}. The gas phase metallicities are estimated using \citet{charlot&longhetti01} models as outlined in \citet{tremonti04}. All measurements of SDSS galaxy physical properties utilised in this work correspond only to the 3\arcsec\ SDSS fiber, which is typically centered on the nuclear region and represents only a fraction of the total galaxy, unless explicitly stated otherwise. As recommended by the MPA/JHU group, we increase the uncertainties associated with each emission line and adopt the values listed in \citet{juneau14}, which are updated for the DR7 dataset. 

For our comparison, we require that all emission lines needed for classification using the Baldwin, Phillips \& Terlevich \citep[BPT;][]{baldwin81} diagram ([OIII]/\hbeta\ vs. [NII]/\halpha) have $S/N\ge5$. We then restrict the sample to galaxies classified as star forming on the BPT diagram and that have a redshift of $z<0.2$. This provides us with a sample of $\sim$156,000 local galaxies for comparison.

\section{Methodology}\label{method}
\subsection{SED Fitting}\label{method_SED_fitting}

\begin{figure*}
\begin{center}
\includegraphics[width=0.98\textwidth]{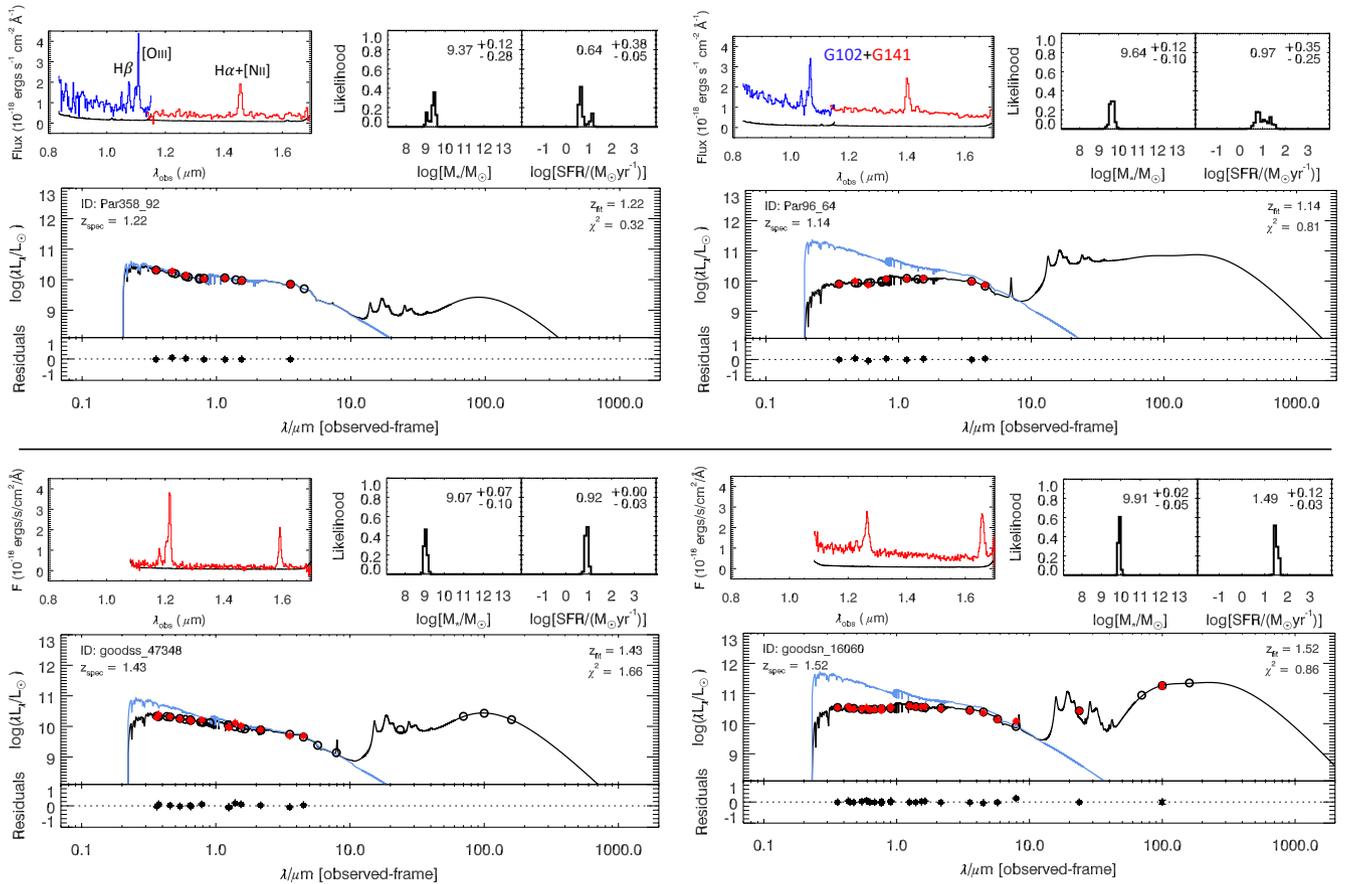}
\end{center}
\vspace*{-0.3cm}
\caption{Examples of \magphys\ best-fit SEDs (black lines) for galaxies with $S/N(\mathrm{H}\beta)>5$ in (\textit{Top}) WISP and (\textit{Bottom}) 3D-HST. The red squares are the observed photometry and the black circles are the corresponding model values. The blue lines show the predicted intrinsic stellar population SEDs (without attenuation). The grism spectra and posterior probability distribution functions (PDFs) for stellar mass and SFR are also shown. SFRs are usually poorly constrained, particularly so for WISP, due to the lack of IR data (26\% of our 3DHST sample has \spitzer\ 24\micron\ and/or Herschel).
 \label{fig:wisp_example}}
\end{figure*}

We perform SED-fitting on our galaxy sample using the \magphys\ (high-$z$) spectral modeling code \citep{daCunha15, battisti20}. The version of \magphys\ used in this study is identical to the version described in \citet{battisti20} and includes a free parameter for the strength of the 2175\ang feature. \magphys\ assumes a parametric star-formation history (SFH), which rises linearly at early ages and then declines exponentially (delayed-tau model) with additional instantaneous bursts of star formation, and a \citet{chabrier03} initial mass function (IMF). This parametrisation could introduce systematic biases to the stellar mass estimates \citep[e.g.,][]{leja19}. Dust attenuation is parametrised using the dust model of \citet{charlot&fall00} for which the interstellar dust is distributed into two components, one associated with star-forming regions (stellar birth clouds) and the other with the diffuse interstellar medium (ISM). The default \magphys\ SFRs are derived from the star formation that occurs over the last 100~Myr timescale, which is roughly consistent with SFR(UV+IR) timescales. This is in contrast to the $\lesssim$10~Myr timescale that is typically associated SFR(\halpha ). To avoid circularity issues, we attempt to minimally rely on SFRs from \magphys\ for our main analysis because of its dependence on the assumed dust attenuation prescription.

\magphys\ does not include templates for emission line fluxes and therefore we perform emission line subtraction prior to SED fitting, when available. This is especially important for this study because we are using an emission-line-selected sample. If one assumes a roughly flat continuum, the average flux density measured in the photometry can be approximated as \citep[e.g.,][]{whitaker14}
\begin{equation}
F_\lambda \simeq F_{\lambda,\mathrm{cont}} + F_\mathrm{line}/\Delta \lambda \,,
\end{equation}
where $F_{\lambda,\mathrm{cont}}$ is the continuum-only flux density and $\Delta \lambda$ is the width of the filter, which we take to be its FWHM. We subtract emission line fluxes for all lines with $S/N>2$ from the photometric data. The impact of emission lines on photometry are the largest for galaxies with fainter continuum emission (typically lower $M_\star$) and higher equivalent widths ($EW=F_\mathrm{line}/F_{\lambda,\mathrm{cont}}$). Examples of \magphys\ fits for two WISP and two 3D-HST galaxies are shown in Figure~\ref{fig:wisp_example}.

For each \magphys\  fit, a goodness-of-fit is determined based on the best-fit model using a reduced $\chi^2$ metric, $\chi_\mathrm{red}^2=\chi^2/N_\mathrm{bands}$, where $N_\mathrm{bands}$ is the number of bands observed with non-zero flux. We exclude cases of poor-quality fits by removing galaxies with $\chi_\mathrm{red}^2>3$ from our final sample. We also require accurate stellar masses, $\sigma(\log M_\star)\leq 0.2$~dex (based on the 16th and 84th percentiles of the posterior PDF; $\sigma(\log M_\star)$=($\log M_{\star\mathrm{,p84}}-\log M_{\star\mathrm{,p16}})/2$), because stellar masses are required for corrections for \halpha+[NII] and stellar absorption. Cases of poor fits are likely to associated with either poor/inconsistent photometric data and/or strong AGN contamination. The median values of the remaining WISP and 3D-HST samples are $\bar{\chi}_\mathrm{red}^2=0.45$ and 0.78. The WISP values are lower than 3D-HST due to the fact that WISP has fewer photometric datapoints and the models can `over-fit' the data (more free parameters than datapoints). For reference, the median 1$\sigma$ uncertainty on $\log M_\star$ is 0.11 and 0.07~dex for WISP and 3D-HST, respectively. The median 1$\sigma$ uncertainty on $\log \mathrm{SFR(SED)}$ is 0.22 and 0.10~dex for WISP and 3D-HST, respectively. Due to the large uncertainty in SFRs when using only UV through near-IR photometry, evident from the uncertainties and also comparing to values in the 3D-HST catalog (see Appendix~\ref{3dhst_compare}), we make minimal use of SFR(SED) for individual galaxies for our main analysis.

\subsection{Characterising Attenuation}\label{method_attenuation}

\subsubsection{Attenuation of Ionised Gas}
The dust attenuation of ionised gas in a galaxy can be quantified through the Balmer decrement, $F(\mathrm{H}\alpha)/F(\mathrm{H}\beta)$, where \halpha\ and \hbeta\ are located at 6562.8\ang and 4861.4\AA , respectively. Following \citet{calzetti94}, we define the Balmer optical depth, $\tau_B^l$, as
\begin{equation}\label{eq:tau}
\tau_B^l = \tau_{\mathrm{H}\beta} - \tau_{\mathrm{H}\alpha} = \ln \left(\frac{F(\mathrm{H}\alpha)/F(\mathrm{H}\beta)}{2.86}\right)\,,
\end{equation}
where $\tau_{\mathrm{H}\beta}$ and $\tau_{\mathrm{H}\alpha}$ are the optical depths at the wavelengths of \hbeta\ and \halpha\, respectively, and the value of 2.86 comes from the theoretical value expected for the unreddened ratio of $F(\mathrm{H}\alpha)/F(\mathrm{H}\beta)$ undergoing Case B recombination with $T_{\mathrm{e}}=10^4$~K and $n_{\mathrm{e}}=100$~cm$^{-3}$ \citep{osterbrock89,osterbrock&ferland06}. The superscript \textit{l} is used to emphasize that this quantity is coming from emission lines and should be distinguished from optical depths associated with the stellar continuum. For most individual galaxies in our sample $\tau_B^l$ cannot be determined due to the non-detection of \hbeta\ and we instead determine average values of a large group of galaxies ($N_\mathrm{stack}\gtrsim100$) based on stacked spectra. 

If one assumes knowledge of the total-to-selective extinction, $k(\lambda)\equiv A_\lambda /E(B-V)$, then $\tau_B^l$ can be directly related to the colour excess of the nebular gas, $E(B-V)_{\mathrm{gas}}$ \citep{calzetti94}, through
\begin{equation}\label{eq:EBV_gas}
E(B-V)_{\mathrm{gas}}=\frac{1.086\tau_B^l}{k(\mathrm{H}\beta)-k(\mathrm{H}\alpha)} \,.
\end{equation}
For reference, the MW, LMC, and SMC have values of $k(\mathrm{H}\beta)-k(\mathrm{H}\alpha)=1.160$, 0.965, and 1.165, respectively \citep{fitzpatrick19, gordon03}. 

\subsubsection{Attenuation of Stellar Continuum}
For galaxies where the UV flux density is dominated by recent star formation (i.e., small contribution from older stars), the intrinsic UV spectral slope for a continuously star forming galaxy has an expected value in the range of $-2.6\lesssim \beta_0\lesssim-2.0$, depending on assumed metallicity, IMF, and binarity of massive stars \citep[e.g.,][]{leitherer99, stanway16}. The effects of dust attenuation are also often inferred from the observed UV spectral slope $\beta$, where
\begin{equation}\label{eq:beta}
F(\lambda)\propto\lambda^\beta\,,
\end{equation}
and $F(\lambda)$ is the flux density in the UV, typically in the range $1250\text{\AA}\le\lambda\le2600\text{\AA}$. 

We determine the UV continuum slope for each galaxy in two ways, described below. The first method is derived directly from the photometric data, $\beta_\mathrm{phot}$ and the second is derived from the best-fit SED from \magphys\ (Section~\ref{method_SED_fitting}), $\beta_\mathrm{SED}$. 

The value of $\beta_\mathrm{phot}$ is determined by fitting the photometry in the range $1250\text{\AA}\le\lambda\le3000\text{\AA}$ according to eq~(\ref{eq:beta}), and requiring at least two data points. We note that this is a slightly longer wavelength range than those commonly adopted for $\beta$ ($1250\text{\AA}\le\lambda\le2600\text{\AA}$), but we justify this in order to minimize the uncertainty on $\beta$ and also the impact that the presence of a 2175\ang feature has on the measured UV-slope, which is particularly problematic with only two bands \citep[e.g.,][]{battisti16, shivaei20a}. This is especially helpful for WISP because it has less rest-frame UV data than 3D-HST. As demonstrated below, this longer wavelength range appears to have minimal bias on the UV-slope relative to the typical measurement uncertainty \textit{for our sample}. As a reference, we find a median change of $\beta_\mathrm{phot}(1250\text{\AA}\le\lambda\le2600\text{\AA})-\beta_\mathrm{phot}(1250\text{\AA}\le\lambda\le3000\text{\AA})=0.05$ and -0.07 for WISP and 3D-HST, respectively. 

The value of $\beta_\mathrm{SED}$ is determined by fitting eq~(\ref{eq:beta}) on the best-fit \magphys\ SED. We use wavelength widows covering the range $1250\text{\AA}\le\lambda\le2600\text{\AA}$ that minimize the effect of stellar absorption features and the 2175\ang feature on the inferred UV slope. These are similar to the definition used in previous studies \citep[e.g.,][]{calzetti94, reddy15}. It is important to note that $\beta_\mathrm{SED}$ is dependent on the assumptions in \magphys\ for the stellar population models \citep{bruzual&charlot03} and the dust attenuation treatment (the two component \citet{charlot&fall00} model, plus a 2175\ang feature of variable strength). 

\begin{figure}
\includegraphics[width=0.45\textwidth]{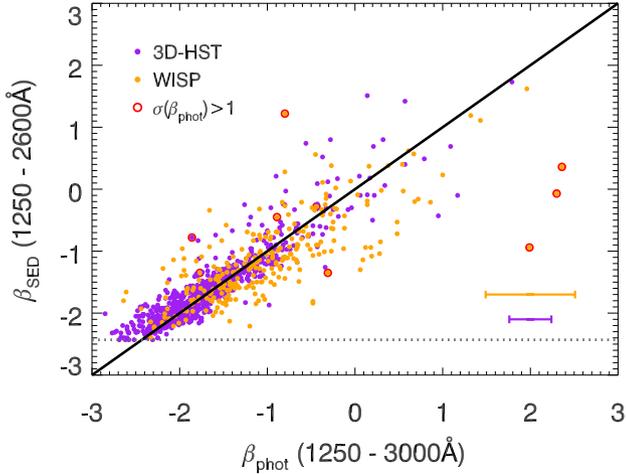}
\caption{Comparison between the photometric-based UV-slopes, $\beta_\mathrm{phot}$, and those from the best-fit \magphys\ SEDs, $\beta_\mathrm{SED}$. The median errorbar on $\beta_\mathrm{phot}$ for each survey is indicated (0.51 for WISP, 0.24 for 3D-HST). There is good agreement, with a median of $\beta_\mathrm{phot}-\beta_\mathrm{SED}=0.10$ and $-0.03$ and $\sigma(\beta_\mathrm{phot}-\beta_\mathrm{SED})=0.59$ and $0.24$ for WISP and 3D-HST, respectively. The dotted horizontal line is the limit of the bluest stellar populations in \magphys\ ($\beta_\mathrm{SED}\sim-2.4$). We exclude galaxies with a UV-slope uncertainty of $\sigma(\beta_\mathrm{phot})>1$ (open red circles) from our analysis.
 \label{fig:UVslope_compare}}
\end{figure}

We compare the UV slopes $\beta_\mathrm{phot}$ and $\beta_\mathrm{SED}$ in Figure~\ref{fig:UVslope_compare}. We find good agreement, in particular for the 3D-HST data which has more ancillary photometry than WISP. For the combined sample, the median offset is $\Delta\beta=\beta_\mathrm{phot}-\beta_\mathrm{SED}=-0.02$ with a scatter of $\sigma(\beta_\mathrm{phot}-\beta_\mathrm{SED})=0.37$. The observed scatter is consistent with the measurement uncertainties. We also compared $\beta_\mathrm{phot}$ to the UV slopes provided in the 3D-HST catalog and found good agreement (see Appendix~\ref{3dhst_compare}). For our main analysis, we impose a cut to exclude galaxies where $\sigma(\beta_\mathrm{phot})>1$ because the derived properties for these galaxies, in particular their SFRs, are likely to be unreliable. To limit potential biases due to the model assumptions in \magphys, we choose to adopt $\beta_\mathrm{phot}$ for our main analysis.

\subsection{Stacking and Emission Line Fitting}\label{method_stacking_fitting}

Galaxy spectra are stacked and fit following similar methods to those described in \citet{henry21}, but modified for stacking based on \halpha\ instead of [OIII]. In brief, we take the continuum-subtracted spectra, determined in the visual inspection/fitting phase, and normalise them by their \halpha\ flux. The spectra are then de-redshifted using a linear interpolation to shift them onto a common grid of rest wavelengths. Finally, we take the median of the normalised fluxes at each wavelength.

To fit the stacked spectra, we fit a set of Gaussian profiles to the lines in the region of $4400\text{\AA} <\lambda_\mathrm{rest}<7100\text{\AA}$, covering the \hbeta , [OIII], \halpha +[NII], and [SII] lines. Similar to \citet{henry21}, we find that single Gaussian components are not sufficient to match the observed line profiles and adopt two Gaussian components for each line, one narrow and one broad component. Multiple components can arise due to kinematic differences among ionising sources (e.g., HII vs AGN), but can also occur in grism spectra due to line profiles having a dependence on the spatial distribution of the emitting sources. The FWHM of the broad component is fixed to be the same for all of the lines and also required to be between 1-4x the FWHM of the narrow components. The amplitudes of the broad components for each line are allowed to vary independently (among positive values).

The emission lines are simultaneously fit with the following assumptions/restrictions: (1) the ratio of [OIII]$\lambda$5007/[OIII]$\lambda$4959 is fixed to 2.98:1 \citep{storey&zeippen00}, (2) single profiles are used for the closely spaced blends of  \halpha +[NII]$\lambda\lambda$6548, 6583, and [SII]$\lambda\lambda$6716, 6731, (3) we require the FWHM of the \hbeta\ and [OIII] narrow components to match and the FWHM of \halpha +[NII], and [SII] narrow components to match separately  (i.e., match close pairs) to account for the spectral resolution difference between the G102 and G141 grisms, (4) we require the FWHM of the narrow components to be within a factor of 2 with each other, (5) we allow a $\pm10\text{\AA}$ shift (rest frame) of the emission line centroids to accommodate systematic uncertainties in the grism wavelength solution, and (6) we account for any (small) residual continuum offsets due to imperfect continuum subtraction by including free parameters for the spectra amplitudes (i.e., constant offsets) in the \hbeta\ and [OIII] region ($4400\text{\AA}<\lambda_\mathrm{rest}<5500\text{\AA}$) and \halpha +[NII] and [SII] region ($6000\text{\AA}<\lambda_\mathrm{rest}<7100\text{\AA}$).

\subsection{Mitigating AGN contamination}\label{method_agn_selection}
For a majority of our sources, we only have limits for emission lines that are commonly utilised for AGN diagnostics. This is further complicated by the low spectral resolution of the \hst\ grism data such that the \halpha\ and [NII] lines are blended. Another common way to identify AGN is through their mid-IR/far-IR colours \citep{donley12, kirkpatrick13}, but only the 3D-HST data has full IRAC coverage to provide identification of AGN candidates in this manner. We identify and exclude 3D-HST AGN candidate sources that satisfy the IRAC colour-colour selection of \citep{donley12}, where we require $S/N>3$ in all four channels. For cases without the required $S/N>3$ in all four channels, we also exclude cases based on the single colour (8.0\micron\ vs 3.6\micron; requiring both $S/N>3$) criteria of \citep{kirkpatrick13}. The \citet{donley12} criteria excludes 10 sources and the \citet{kirkpatrick13} criteria excludes an additional 9 sources. We do not use X-ray emission to identify AGN candidates because X-ray data are not available for the WISP fields.

\begin{figure}
\begin{center}
\includegraphics[width=0.45\textwidth]{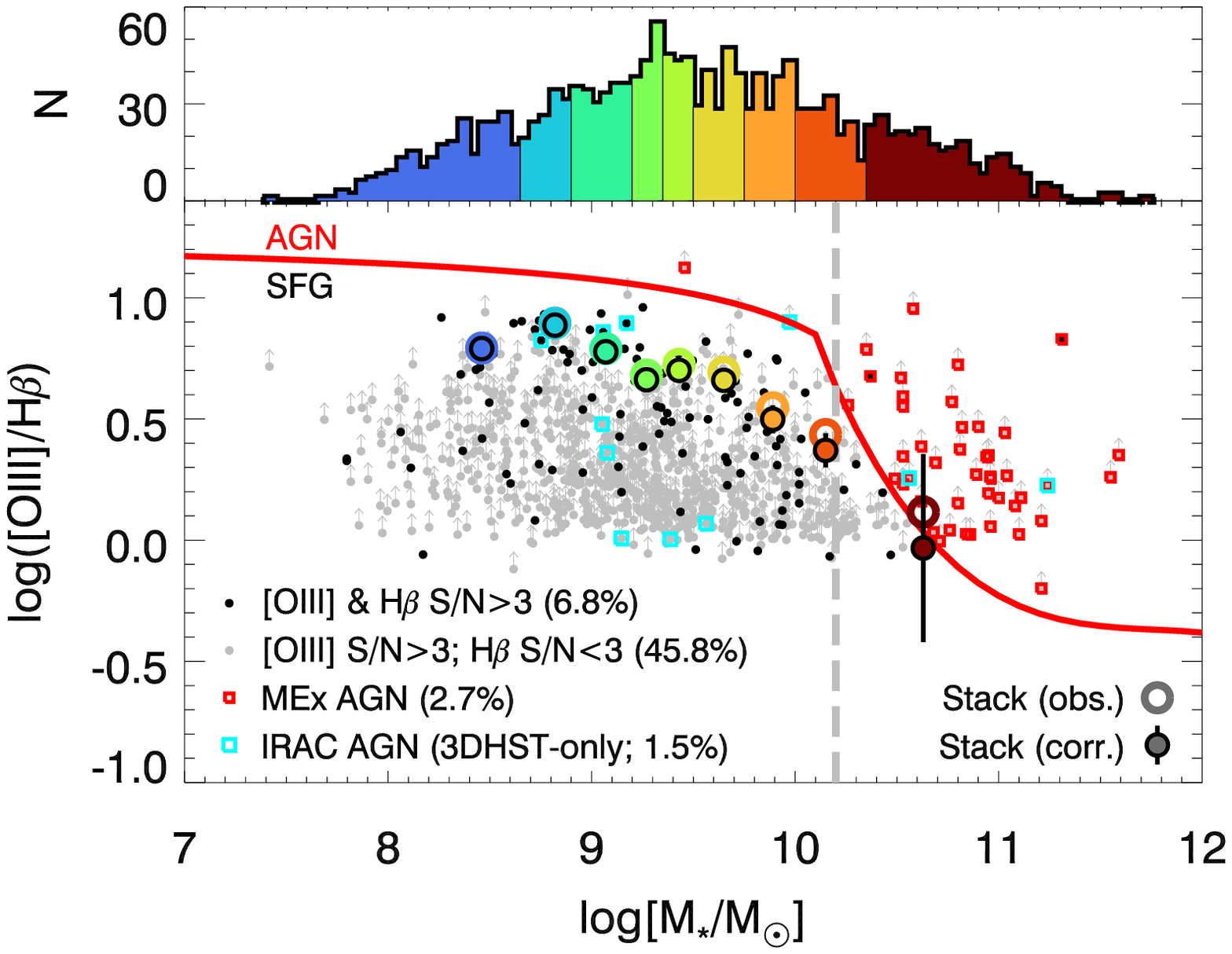}
\end{center}
\vspace*{-0.3cm}
\caption{Mass-Excitation (MEx) diagram for galaxies in our sample (Section~\ref{selection}). The red line corresponds to the lower boundary of \citet{juneau14} shifted to higher masses by 0.5~dex (i.e., to the right) to account for redshift evolution \citep{kashino19} and up by 0.13~dex to account for using total [OIII] instead of only [OIII]$\lambda$5007. AGN identified using the MEx and IRAC colour-colour diagnostics are indicated by small red and cyan open squares, respectively. 
The large coloured circles correspond values from spectra stacked according to galaxy stellar mass (see Section~\ref{result_stack}), but excluding all AGN candidates, with open and filled symbols denoted values before and after making stellar absorption corrections (Section~\ref{method_blend_abs}). \hbeta\ lines for individual galaxies are \textit{not} corrected for stellar absorption. The large uncertainty in the highest mass stack is due to a combination of large measurement uncertainty (weak lines) and uncertainty in the \hbeta\ stellar absorption correction. Stacks below $\log (M_\star /M_\odot)=10.2$ (vertical dashed gray line) lie below the AGN line, regardless of corrections, suggesting they are not dominated by AGN contamination. 
 \label{fig:MEx}}
\end{figure}

Another technique to identify AGN candidates is based on the ratio of [OIII]/\hbeta\ vs. stellar mass, known as the Mass-Excitation (MEx) diagram \citep{juneau14}. It has been demonstrated that the demarcation line between star forming galaxies (SFGs) and AGN in this diagram appears to evolve with redshift \citep{coil15, kashino19}. Using a sample of $z\sim1.6$ galaxies, \citet{kashino19} found that the best demarcation line is shifted from the $z\sim0$ relation of \citet{juneau14} by $\Delta \log M_\star=+0.5$. As our sample ($z\sim1.3$) is at roughly comparable redshifts to the sample in \citet{kashino19}, we choose to adopt the same shift to identify AGN candidates in the MEx diagram ($\Delta \log M_\star=+0.5$). We also apply a vertical offset of 0.13~dex due to our [OIII] grism measurements being total values ([OIII]=[OIII]$\lambda$5007+[OIII]$\lambda$4959; we assume a ratio of 2.98:1, respectively), whereas the original MEx diagram uses only [OIII]$\lambda$5007. The MEx diagram for our sample is shown in Figure~\ref{fig:MEx}. Unfortunately, only a small fraction (6.8\%) of the galaxies in our sample are detected in both [OIII] and \hbeta , 45.8\% of cases are detected in [OIII] but not \hbeta , and roughly half of the sample (47.4\%) is undetected in both [OIII] and \hbeta\ and therefore cannot be classified using the MEx diagram on an individual level. We identify and exclude 25 and 20 galaxies as candidate AGN based on the MEx criteria in WISP and 3D-HST, respectively. For 3D-HST, 37 galaxies in total are excluded based on IRAC colours and the MEx diagram (2 sources are identified in both).

Based on stacking results presented in Section~\ref{result_stack}, we also make a conservative cut to exclude galaxies at $\log (M_\star /M_\odot)>10.2$ for our main analysis because they may have a non-negligible AGN fraction in the sample. Below this mass, the emission line ratios of our stacks do not appear to be dominated by AGN as they remain below the AGN line in the MEx diagram. For reference, \citet{forsterSchreiber19} find that galaxies with $\log (M_\star /M_\odot)<10.2$ have an AGN occurrence rate of $f_{AGN}\lesssim10$\%, based on a sample of 600 galaxies at redshift $0.6<z<2.7$, with $f_{AGN}$ increasing dramatically with increasing mass (e.g., $\sim$60\% at $\log (M_\star /M_\odot)=11$; see their Figure~6).

\subsection{Corrections for \halpha +[NII] Blending and Stellar Absorption}\label{method_blend_abs}
We correct for the blending of \halpha +[NII] using the stellar mass- and redshift-dependent functional relation from \citet{faisst18},  which used $\sim$190,000 SDSS galaxies combined with the observed BPT locus evolution of SFGs from $0<z<2.5$. The corrections are stated to be valid over the ranges of $0<z<2.7$ and $8.5 < \log(M_\star/ M_\odot) < 11.0$, with an intrinsic scatter of $\sim$0.2~dex. A small fraction of our stellar masses extend below $\log(M_\star/ M_\odot)=8.5$ but extrapolating the \citet{faisst18} relation to lower masses should have minimal impact as the expected [NII] contribution is $\lesssim5\%$ at all redshifts in this mass regime. Stellar masses come from the median of the posterior probability distribution functions (PDFs) from the \magphys\ SED fits. If we apply these corrections to our sample (excluding AGN), we find that the median fraction of flux that is in the [NII] line is 12\%. For our main analysis, we minimize the impact of intrinsic variation in this ratio by \textit{only applying blending corrections on stacked spectra}, which are based on groups of $N>100$ galaxies. We note that \citet{martens19} derived a functional form for the [NII]/\halpha\ ratios at $1.3\leq z <2.1$ that differs from that of \citet{faisst18} at high stellar masses ($\log(M_\star/ M_\odot)>10.2$) by up to a factor of 2, indicating that the corrections in the high-mass regime are less certain at these redshifts.

Another effect that will alter Balmer decrement measurements is stellar absorption, which cannot be directly fit in the low resolution grism spectra. To correct for stellar absorption, we use the stellar mass- and SFR-dependent functional relation from \citet{kashino&inoue19}, which is based on trends observed for $\sim$190,000 SDSS galaxies. The fractional corrections were determined over the range of $7.2 < \log(M_\star/ M_\odot) < 11.4$, with an intrinsic scatter of $\sim$10-20\%. Galaxies with lower sSFR (i.e., older average stellar populations) require larger fractional corrections for stellar absorption than those with higher sSFR.
Similar to the \halpha +[NII] blending corrections, we attempt to minimize the impact of intrinsic variation by \textit{only applying Balmer absorption corrections on stacked spectra for our main analysis}. For each stacked bin, we determine the median value of stellar masses and SFRs from \magphys. However, if the inferred SFR from \magphys\ is lower than the SFR inferred from the median observed \halpha\ for that bin (corrected for [NII] blending), then we adopt the latter as the input to determine the fraction of stellar absorption because it should be a lower-limit. 
For reference, if we apply corrections to individual galaxies we find the median value of the inferred absorbed fractions for \halpha\ and \hbeta\ in our sample is 1.7\% and 5.3\%, respectively. These low values are driven by the relatively high sSFR of our ELGs (see Section~\ref{result_MS}).

\section{Results}\label{results}
\subsection{Stacked Spectra based on Stellar Mass}\label{result_stack}
\begin{figure*}
\begin{center}
\includegraphics[width=0.95\textwidth]{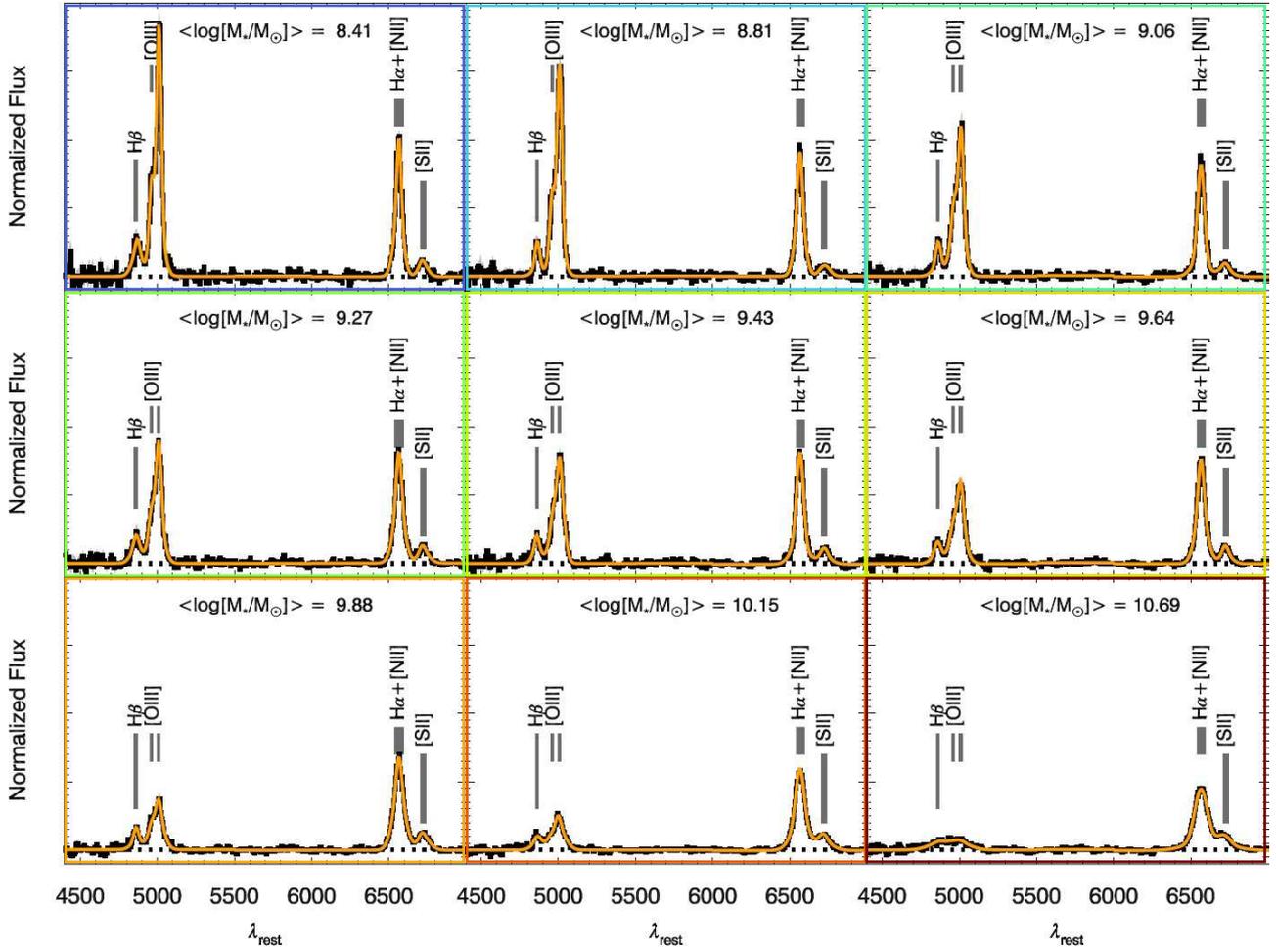}
\end{center}
\vspace{-0.3cm}
\caption{Stacked spectra normalised to \halpha+[NII] total flux in bins of stellar mass for our sample (Section~\ref{selection}) and excluding AGN candidates (Section~\ref{method_agn_selection}). The average stellar mass value is indicated at the top of each panel. The spectral fits (Section~\ref{method_stacking_fitting}) are shown as the orange line. The colours outlining the panels coincide with symbol colours in figures that use the mass stack results (Figures~\ref{fig:MEx}, \ref{fig:UV_slope_vs_mass}, and \ref{fig:galaxy_MS}).  
\label{fig:mass_stacks}}
\end{figure*}

As a first examination of the data, we bin the sample according to stellar mass, $M_\star$. Stellar mass is a relatively robust physical parameter that is minimally affected by dust attenuation \citep[$\Delta \log M_\star\leq0.3$~dex; e.g.,][]{conroy13}. Using the methods described in Section~\ref{method_stacking_fitting}, we generate stacked spectra for 9 equal-number stellar mass bins (each $N\sim118$ galaxies) and fit their emission lines. These stacks and their fits are shown in Figure~\ref{fig:mass_stacks}. The emission line ratios and average properties of the stacks are shown in Table~\ref{tab:stack_mass}. We note that the line measurements for \hbeta\ and [OIII] for the highest mass bin are poorly constrained because they are less well defined (relative to other stacks) and appear quite broad (\hbeta\ and [OIII] blend together), potentially due to AGN contamination (discussed below), which may introduce biases in the fitted values.

\setlength{\tabcolsep}{3pt}
\begin{table*}
\caption{Summary of average properties and line ratios for WISP+3DHST stacked spectra binned by stellar mass (full mass range) \label{tab:stack_mass}} \vspace{-0.2cm}
\begin{center}
\begin{tabular}{ccccccccccccc}
 \hline \hline
\vspace{0.02in} \multirow{2}{*}{$N$} & 
\multirow{2}{*}{$\left\langle \log M_\star \right\rangle$} & 
\multicolumn{2}{c}{$\left\langle \log \mathrm{SFR} \right\rangle$} & 
\multirow{2}{*}{$\left\langle \beta_\mathrm{phot} \right\rangle$} & 
\multirow{2}{*}{$\dfrac{\mathrm{[OIII]}}{\mathrm{H}\beta}$}  & 
\multirow{2}{*}{$\dfrac{\mathrm{[OIII]}}{\mathrm{H}\beta_\mathrm{corr}}$} & 
\multirow{2}{*}{$\dfrac{(\mathrm{H}\alpha\mathrm{+[NII]})}{\mathrm{[OIII]}}$} & 
\multirow{2}{*}{$\dfrac{(\mathrm{H}\alpha\mathrm{+[NII]})}{\mathrm{H}\beta}$} & 
\multirow{2}{*}{$\dfrac{\mathrm{H}\alpha}{\mathrm{H}\beta}$} & 
\multirow{2}{*}{$\dfrac{\mathrm{H}\alpha_\mathrm{corr}}{\mathrm{H}\beta_\mathrm{corr}}$}& 
\multirow{2}{*}{$\tau_B^l$} & 
\multirow{2}{*}{$\dfrac{\mathrm{[SII]}}{(\mathrm{H}\alpha\mathrm{+[NII]})}$} \\ 
 & & $\mathrm{H}\alpha_\mathrm{corr}$ & SED  \\
 \hline
     119 &  8.41 &  0.36 &  0.19 & -2.13 &  6.20 $\pm$ 0.54 &  6.17 $\pm$ 0.54 &  0.50 $\pm$ 0.01 &  3.08 $\pm$ 0.27 &  2.93
 $\pm$ 0.26 &  2.92 $\pm$ 0.26 &  0.02 $\pm$ 0.09 &  0.15 $\pm$ 0.01 \\
     119 &  8.81 &  0.67 &  0.36 & -1.90 &  7.86 $\pm$ 0.59 &  7.71 $\pm$ 0.60 &  0.57 $\pm$ 0.01 &  4.49 $\pm$ 0.33 &  4.18
 $\pm$ 0.31 &  4.14 $\pm$ 0.31 &  0.37 $\pm$ 0.07 &  0.12 $\pm$ 0.01 \\
     119 &  9.06 &  0.62 &  0.52 & -1.82 &  6.17 $\pm$ 0.54 &  5.99 $\pm$ 0.55 &  0.65 $\pm$ 0.02 &  4.04 $\pm$ 0.35 &  3.69
 $\pm$ 0.32 &  3.63 $\pm$ 0.32 &  0.24 $\pm$ 0.09 &  0.17 $\pm$ 0.01 \\
     118 &  9.27 &  0.68 &  0.60 & -1.71 &  4.80 $\pm$ 0.41 &  4.59 $\pm$ 0.44 &  0.84 $\pm$ 0.02 &  4.02 $\pm$ 0.34 &  3.60
 $\pm$ 0.31 &  3.50 $\pm$ 0.32 &  0.20 $\pm$ 0.09 &  0.18 $\pm$ 0.01 \\
     118 &  9.43 &  0.89 &  0.70 & -1.59 &  5.32 $\pm$ 0.35 &  5.02 $\pm$ 0.41 &  0.94 $\pm$ 0.02 &  4.99 $\pm$ 0.32 &  4.38
 $\pm$ 0.28 &  4.21 $\pm$ 0.32 &  0.39 $\pm$ 0.08 &  0.18 $\pm$ 0.01 \\
     118 &  9.64 &  0.95 &  0.84 & -1.54 &  4.94 $\pm$ 0.35 &  4.57 $\pm$ 0.42 &  1.06 $\pm$ 0.03 &  5.23 $\pm$ 0.36 &  4.44
 $\pm$ 0.31 &  4.20 $\pm$ 0.35 &  0.38 $\pm$ 0.08 &  0.18 $\pm$ 0.01 \\
     118 &  9.88 &  1.10 &  0.97 & -1.35 &  3.52 $\pm$ 0.32 &  3.15 $\pm$ 0.42 &  1.67 $\pm$ 0.06 &  5.88 $\pm$ 0.50 &  4.72
 $\pm$ 0.43 &  4.35 $\pm$ 0.52 &  0.42 $\pm$ 0.12 &  0.21 $\pm$ 0.01 \\
     118 & 10.15 &  1.15 &  1.15 & -1.05 &  2.73 $\pm$ 0.19 &  2.35 $\pm$ 0.38 &  2.42 $\pm$ 0.08 &  6.59 $\pm$ 0.41 &  4.74
 $\pm$ 0.37 &  4.23 $\pm$ 0.61 &  0.39 $\pm$ 0.14 &  0.21 $\pm$ 0.01 \\
     118 & 10.69 &  0.87 &  1.35 & -0.55 &  1.31 $\pm$ 0.51 &  0.93 $\pm$ 0.83 &  5.13 $\pm$ 1.22 &  6.70 $\pm$ 2.09 &  4.10
 $\pm$ 1.29 &  3.11 $\pm$ 1.84 &  0.08 $\pm$ 0.59 &  0.22 $\pm$ 0.01 \\
\hline
 \end{tabular}
\end{center}
 \textbf{Notes.} Uncertainties on the [NII] blending and stellar absorption corrections (latter denoted by $X_\mathrm{corr}$) are determined as the $1\sigma$ dispersion in correction values for individual galaxies in the bin. These are added in quadrature with the line measurement uncertainties. We stress that this \textit{does not} take the systematic uncertainties in the corrections into account (see Section~\ref{method_blend_abs}). The measurement uncertainties are typically dominant over the correction `uncertainties', except for the highest mass bin where the stellar absorption corrections on \hbeta\ are very significant.
\end{table*}

A few notable trends are evident in the stellar mass-stacked spectra. First, the width of the emission lines tends to increase with increasing stellar mass. It is easiest to see this for \halpha+[NII], because the spectra are normalised to this line. Linewidths in grism data reflect both spatial and kinematic information and this is likely a consequence of both an increase in size (spatial extent) of galaxies and an increase in rotational velocity \citep[e.g., Tully-Fisher relation;][]{tully&fisher77} with increasing stellar mass. Second, the strength of the [OIII] emission line relative to \halpha+[NII] decreases with increasing mass. This appears to be correlated with dust attenuation (Section~\ref{other_stacks}), but also note that the dominant [OIII] cooling line is expected vary with ISM conditions (electron temperature, density, metallicity) with the [OIII]$\lambda$88\micron\ emission line tending to becoming dominant at higher metallicities \citep[higher stellar masses; e.g.,][]{stasinska02}. Last, the strength of the [SII] doublet remains roughly similar relative to the \halpha+[NII] line with varying stellar mass.

\begin{figure}
\includegraphics[width=0.45\textwidth]{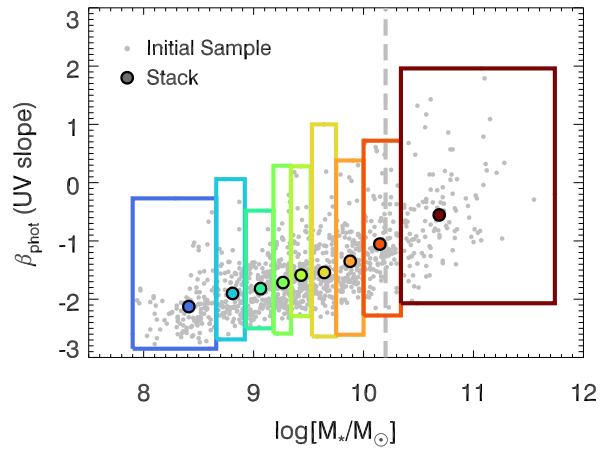}
\caption{The observed UV slope, $\beta_\mathrm{phot}$, as a function of stellar mass. Coloured symbols denote the medians from stellar mass stacks. Gray points show our sample of galaxies (criteria (1)-(3), removing AGN candidates, poor-quality SED fits). A trend is evident that more massive galaxies have redder UV colours (larger $\beta_\mathrm{phot}$). The vertical long-dash gray line corresponds to $\log (M_\star /M_\odot)=10.2$, above which we exclude from our attenuation analysis because stacked spectra in this range may be contaminated by AGN (Section~\ref{method_agn_selection}) and their corrections for [NII] blending and stellar absorption are more uncertain (Section~\ref{method_blend_abs}.
 \label{fig:UV_slope_vs_mass}}
\end{figure}

A plot of the observed UV slope, $\beta_\mathrm{phot}$, as a function of the stellar mass is shown in Figure~\ref{fig:UV_slope_vs_mass}.  There is an obvious correlation where more massive galaxies have a larger $\beta_\mathrm{phot}$, corresponding to redder UV colour. This is a consequence more massive galaxies having both older average stellar population ages (lower sSFR) and more dust attenuation, as both of these act to make the UV slope redder in colour. In contrast, the mass stacks do not show a clear monotonic trend in $\tau_B^l$ (dust attenuation) despite a strong trend in UV reddening (see Table~\ref{tab:stack_mass}). We attribute this difference to a combination of (1) the uncertainties associated with [NII] line blending and Balmer absorption corrections (affecting $\tau_B^l$; see Section~\ref{method_blend_abs}), (2) the effect of differing average stellar ages on $\beta_\mathrm{phot}$, and/or (3) differences in the average star-dust geometry as a function of stellar mass.

We show the mass and [OIII]/\hbeta\ values of these stacks on the MEx diagram in Figure~\ref{fig:MEx}. This allows us to assess the issue of AGN contamination for our entire population of galaxies, including those with $S/N<3$ in both [OIII] and \hbeta\ ($\sim$half of sample). Based on the observed line ratios of the stacks, we use $\log (M_\star /M_\odot)=10.2$ as the threshold above which AGN contamination may significantly affect our stacking results. In addition, two additional factors that impact our line measurements at higher masses are that the \hbeta\ emission line is faint for the highest mass bin and deblending corrections for \halpha+[NII] and \hbeta\ stellar absorption become more uncertain (Section~\ref{method_blend_abs}; see $\tau_B^l$ uncertainty values listed in Table~\ref{tab:stack_mass}). Together, these factors led to our decision to exclude galaxies with $\log (M_\star /M_\odot)>10.2$ from our main dust attenuation analysis because the Balmer decrement measurements are significantly less reliable. It is important to note that this selection introduces a bias in our sample against the most strongly attenuated systems (i.e., reddest UV colours; see Figure~\ref{fig:UV_slope_vs_mass}), which can may affect the derived attenuation curve. This is because galaxies with higher dust attenuation show a preference for shallower dust attenuation curves \citep[e.g., see Figure~10 in][]{salim&narayanan20}. Deeper grism surveys where both [OIII] and \hbeta\ can be individually detected in galaxies for AGN classification would provide the ability to circumvent this issue.

\subsubsection{Sample comparison to the Galaxy Main Sequence}\label{result_MS}
\begin{figure}
\begin{center}
\includegraphics[width=0.45\textwidth]{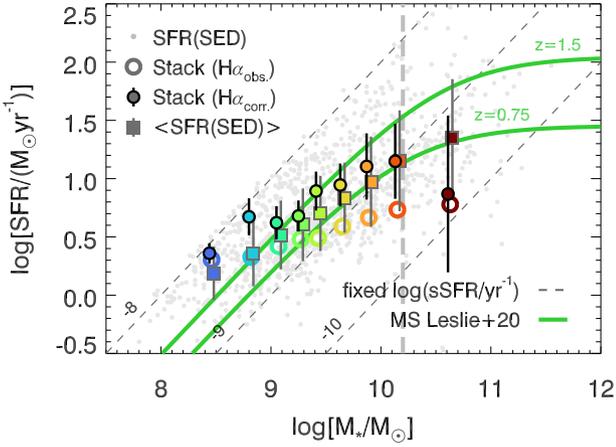}
\end{center}
\vspace{-0.3cm}
\caption{The galaxy main-sequence (MS; logSFR vs. log$M_\star$), for our sample binned by stellar mass (large coloured symbols, offset for clarity). Bins are based on stacked \halpha\ without and with dust corrections from the Balmer decrement, $\mathrm{SFR(H}\alpha_\mathrm{obs})$ and $\mathrm{SFR(H}\alpha_\mathrm{corr})$, respectively, and the median value from SED fitting, $\left\langle\mathrm{SFR(SED})\right\rangle$.
Values for individual galaxies, based on SFR(SED), are shown for reference, with a boundary occurring at $\mathrm{log(sSFR/yr}^{-1})\sim-8$ due to the timescale adopted for SFRs ($10^8$~yr). 
The solid green lines are the galaxy MS from \citet{leslie20} at $z=0.75$ and $z=1.5$ (sample boundaries). Our sample mostly coincides with the MS except for a bias at low masses, likely due to the \halpha\ detection threshold (Section~\ref{result_MS}). We note that the highest-mass bin has very large uncertainty in $\mathrm{SFR(H}\alpha_\mathrm{corr})$ due to large uncertainty in $\tau_B^l$ and that galaxies at $\log (M_\star /M_\odot)>10.2$ (vertical long-dash gray line) are considered less reliable (see Section~\ref{result_stack}). 
 \label{fig:galaxy_MS}}
\end{figure}

To examine how representative our sample is of typical SFGs at these redshifts, we can compare our sample with respect to the galaxy Main Sequence (MS), the relationship between stellar mass and dust-corrected SFR \citep[e.g.,][]{brinchmann04, speagle14, leslie20}. Due to the fact that  \hbeta\ is not detected for a majority of individual galaxies in our sample, we cannot determine the dust-corrected SFR of individual galaxies without assumptions on the dust attenuation curve. A comparison of our stacked spectra based on stellar mass, both before and after dust correction using the Balmer decrement, to the galaxy MS at these redshifts from \citet{leslie20} is shown in Figure~\ref{fig:galaxy_MS}. The \citet{leslie20} relations are based on stacked radio data from $\sim$200,000 galaxies in the COSMOS field. For our SFRs, we adopt the conversion from \cite{kennicutt&evans12}:
\begin{equation}
\log \mathrm{SFR(H}\alpha)=\log L(\mathrm{H}\alpha)-41.27 \,
\end{equation}
which assumes the IMF of \citet[][this is comparable to IMF used in \magphys]{kroupa&weidner03}. For reference, we also show individual SFRs based on the MAGPHYS SED fits, $\mathrm{SFR(SED})$, but note that these are not adopted for our main analysis because of their dependence on the assumed dust attenuation curve (Section~\ref{method_SED_fitting}).  Based on $\mathrm{SFR(SED})$, 98.2\% of our sample lies within 1~dex of the MS at their respective redshifts, with nearly all outliers occurring at $\log (M_\star /M_\odot)>10.2$.

There are several notable aspects that can be highlighted. First, there is a bias at low masses due to the emission line sensitivity selection effect. These limits correspond roughly to $\log \mathrm{SFR_\mathrm{limit}(H}\alpha_\mathrm{obs})\sim -0.2$ and 0.6 at the lower and upper redshift boundary of WISP, respectively, and $\log \mathrm{SFR_\mathrm{limit}(H}\alpha_\mathrm{obs})\sim 0.1$ and 0.2 at the lower and upper redshift boundary range of 3D-HST. At low stellar masses ($\log M_\star \lesssim 9$), galaxies tend to reside above the MS and at $\log M_\star \gtrsim 9$ they reside on the MS, with the exception of our highest mass bin being below the MS. We note that the Balmer decrement of our second mass bin seems anomalously high, leading to a large inferred dust correction, with respect to adjacent bins and given that the average UV slope for this bin is quite blue ($\left\langle\beta_\mathrm{phot}\right\rangle=-1.9$). We explored possible reasons for this by reinspecting the grism spectra for this bin, but did not find any obvious reasons for this effect. It does appear that the \hbeta\ profile in the stacked spectra for this bin is narrower than other bins (Figure~\ref{fig:mass_stacks}), leading to lower \hbeta\ flux. It is also possible that this could be due to AGN contamination or simply reflect a limitation in the accuracy to which the continuum shape can be fit and subtracted. For our main dust attenuation analysis, we use coarser bin sampling (larger number of galaxies per bin) such that this issue should be mitigated. Finally, it can be seen that, with the exception of the last bin, the values of $\mathrm{SFR(H}\alpha_\mathrm{corr})$ are higher than SED-inferred SFR, $\left\langle\mathrm{SFR(SED})\right\rangle$, although most are consistent within their 1$\sigma$ errors. We attribute this effect mainly to the fact that minimal information to constrain dust attenuation is available from the photometry alone for most galaxies (WISP does not have IR data and $\sim$3/4 of the 3D-HST sample is undetected in the available IR data), but also note that differences in SFR indicators may also play a role \citep[e.g.,][]{mcquinn15}. In the absence of IR data, the age-dust degeneracy (older populations can produce redder colours) will result in older (lower SFR) templates in \magphys\ being able to reproduce the data. Including the emission line information in the SED fitting procedure could remove this issue but is beyond the scope of this paper. We also note that the attenuation prescription adopted in \magphys\ could also lead to systematic offsets between these two values.

\subsection{Finding good proxies for the Balmer decrement}\label{other_stacks}

Due to the fact that the \hbeta\ emission line is undetected in the majority of our ELGs ($\sim$90\%), we must rely on using stacked spectra to measure Balmer decrements for $\tau_B^l$. Therefore, we need to find an alternative line ratio or galaxy property that can act as a good proxy for Balmer decrements such that it can be used to sort our sample into bins that monotonically relate to $\tau_B^l$ (i.e., dust attenuation). The final goal is to use these bins to derive average dust attenuation curves. We only consider lines and line ratios that combine \halpha\ +[NII] (or only \halpha) and [OIII] because these are typically the brightest lines available in our grism sample. For reference, our sample has 65.6\% of galaxies with $S/N(\mathrm{[OIII]}) \geq 3$ and 84.4\% with $S/N(\mathrm{[OIII]}) \geq 2$ (and by design, 100\% have $S/N(\mathrm{H}\alpha+\mathrm{[NII]}) \geq 3$). 

\subsubsection{Benchmark from SDSS ($z<0.2$)}
To explore suitable proxies for dust attenuation, we first examine the strength of correlations between various properties with $\tau_B^l$ in the local galaxies from SDSS (defined in Section~\ref{data_sdss}). The corresponding Spearman's rank correlation coefficient, $\rho$, for each galaxy property or line measurement with $\tau_B^l$ in SDSS are summarised in Table~\ref{tab:Spearman_summary}. 

\begin{figure*}
\begin{center}
$\begin{array}{ccc}
\includegraphics[scale=0.58]{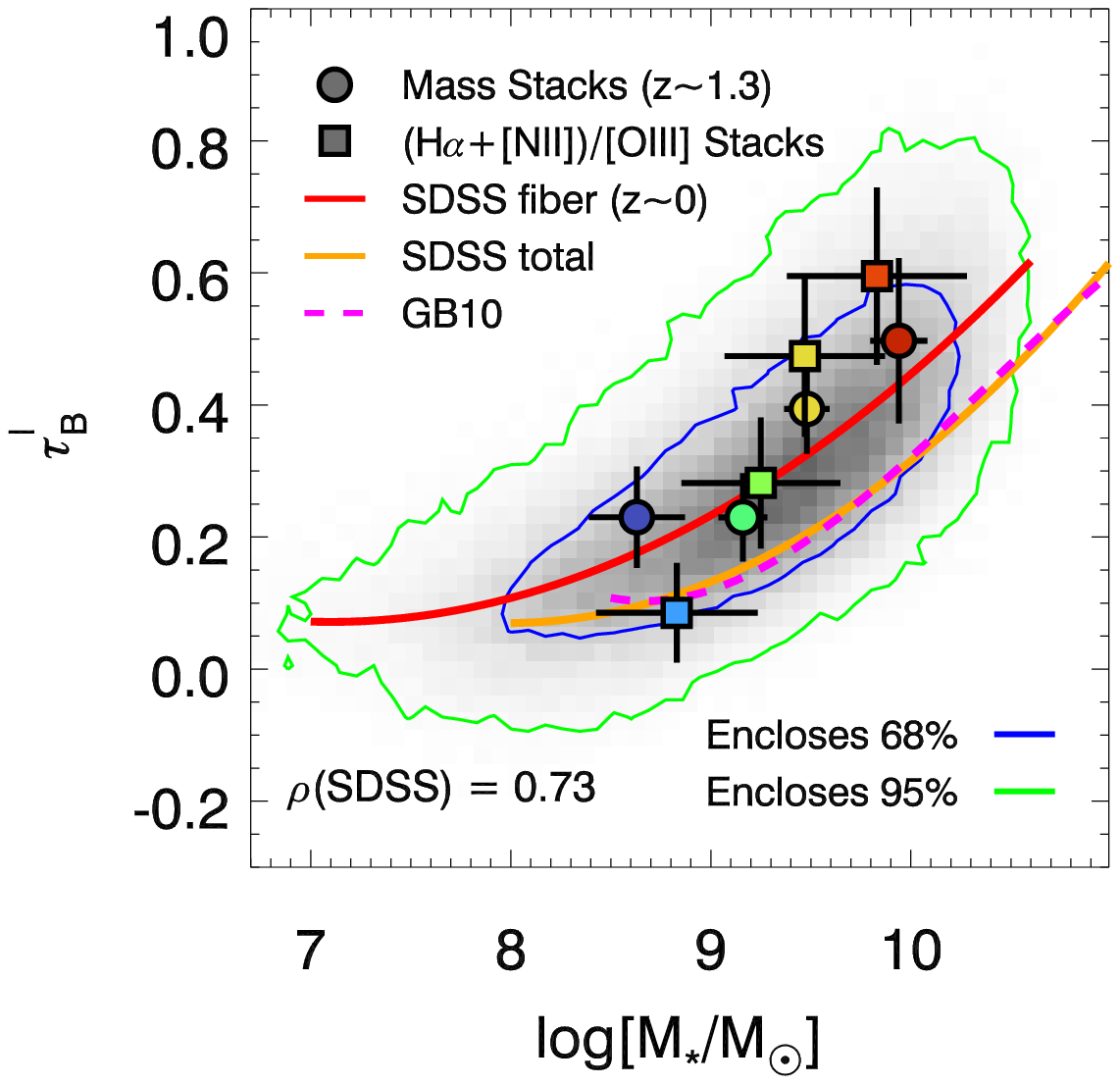} &
\includegraphics[scale=0.58]{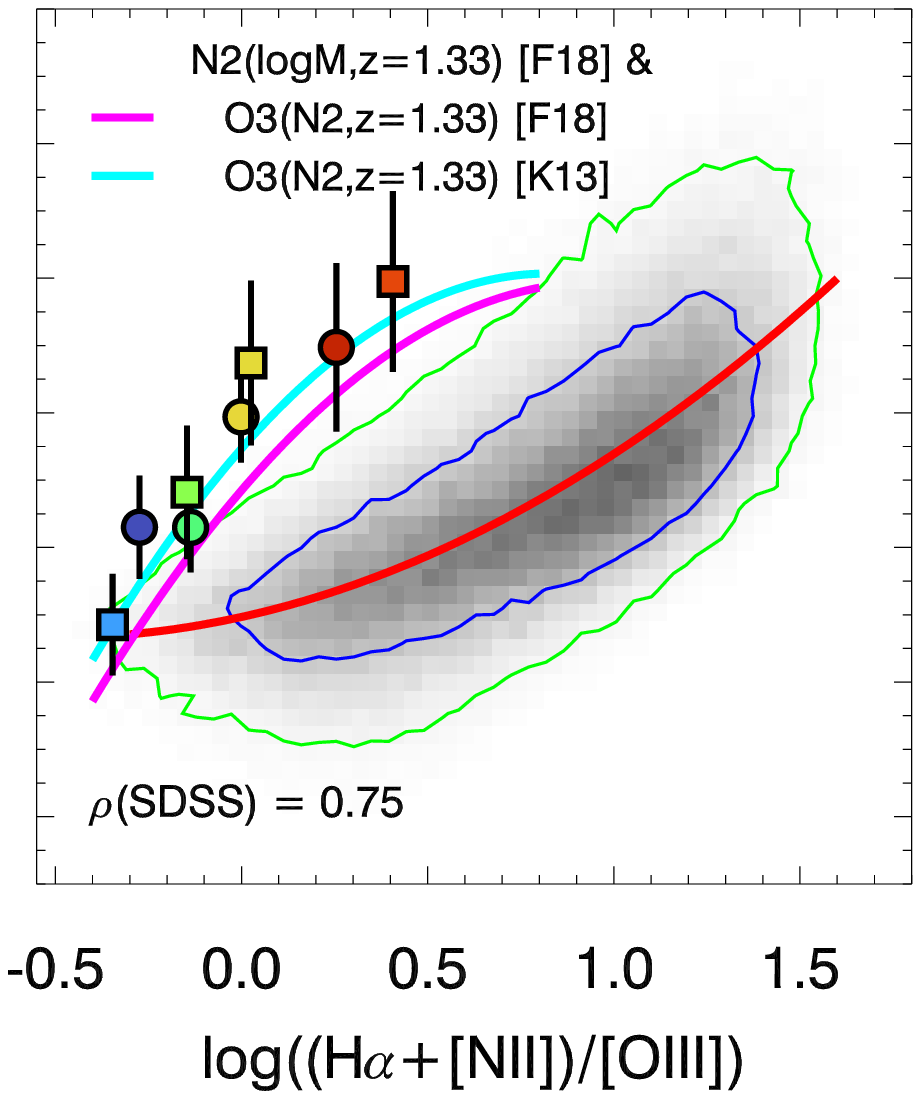} &
\includegraphics[scale=0.58]{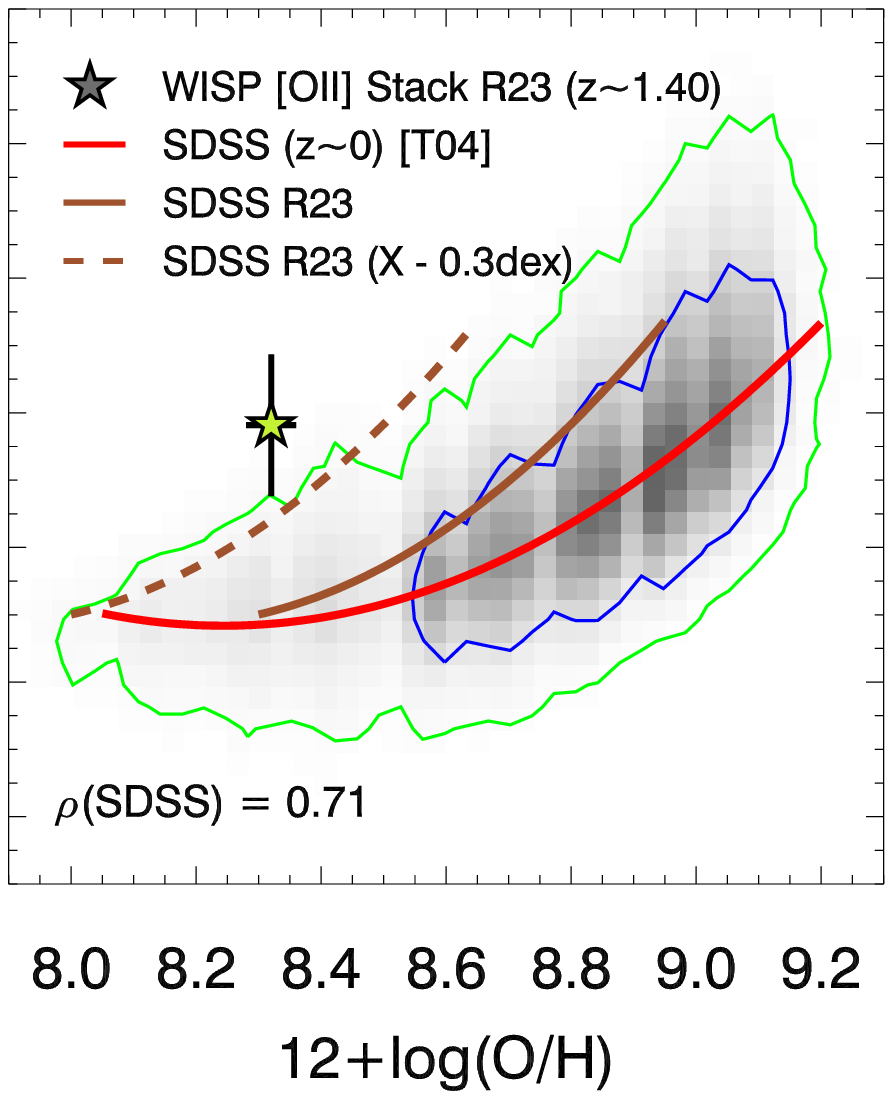} \\
\end{array}$
\end{center}
\vspace{-0.3cm}
\caption{The grayscale density histograms show the Balmer optical depth, $\tau_B^l$, as a function of fiber-region properties for ELGs in SDSS at $z<0.2$, with fits shown by red solid lines. The values from WISP+3DHST stacks are overplotted (only single bin for metallicity; based on R23), where symbol colours relate to the median stellar mass of the bin (similar to Figure~\ref{fig:MEx}). \textit{(Left:)} Stellar mass ($\log M_\star$). We also show the relation if total stellar masses are used for SDSS ($\log M_{\star,\mathrm{tot}}$; orange solid line), which is offset by $\log M_{\star,\mathrm{tot}}\sim \log M_\star+0.5$, and identical to the relation from \citet[][pink dashed line]{garn&best10}.  
\textit{(Middle:)} log((\halpha+[NII])/[OIII]). We show the predicted line ratio change for SDSS galaxies at $z=1.33$ ($\tau_B^l$ and $\log M_\star$ fixed) based on the relations for N2=[NII]/\halpha\ ratio and the BPT SFG locus (O3=[OIII]/\hbeta\ vs. N2) from \citet{faisst18} and \citet{kewley13a}. 
\textit{(Right:)} Gas-phase metallicity \citep[from][]{tremonti04}. The local SDSS relation using the R23 diagnostic \citep{curti17} is also shown, as well offset by $-0.3$~dex lower metallicity ($\tau_B^l$ and $\log M_\star$ fixed). 
The SDSS and WISP+3DHST results are similar for the $\tau_B^l$-$\log M_\star$ relation (fiber) but significantly different for log((\halpha+[NII])/[OIII]) and metallicity. These differences are roughly consistent with expectations from evolving ISM properties with redshift (discussed in Section~\ref{proxy_compare}). 
 \label{fig:tau_property_correlations}}
\end{figure*}

We find that the strongest correlations ($\rho>0.7$) with $\tau_B^l$ are for log((\halpha+[NII])/[OIII]), (or log(\halpha/[OIII])), $\log M_\star$, and 12+log(O/H). Interestingly, we find that the correlation when the [NII] line is not combined with \halpha\ in the ratio with [OIII] is not stronger (in fact it is slightly weaker), suggesting that the line blending does not negatively affect the correlation. A moderate correlation is found between $\tau_B^l$ and SFR, but it is not an ideal proxy to use because SFRs depend on assumptions for dust corrections. We do not find significant correlations with the (observed) flux of \halpha, \halpha +[NII], or [OIII],  the equivalent width of \halpha\ (EW(\halpha)), or sSFR. We also highlight the correlation value found in \citet{battisti16} between the UV slope, $\beta_\mathrm{phot}$ and $\tau_B^l$ based on a sample of $\sim$10,000 galaxies in SDSS that were matched to UV data from \textit{GALEX}, which is weaker than those relative to log((\halpha+[NII])/[OIII]), $\log M_\star$, and 12+log(O/H). The poor correlation between $\tau_B^l$ and the UV slope is most likely linked to the fact that the intrinsic UV slope of galaxies vary with SFH \citep[e.g.,][]{calzetti21} and that this relation also depends on the differential reddening factor between the ionised gas and stellar components \citep[$E(B-V)_{\mathrm{star}}/E(B-V)_{\mathrm{gas}}$; see][for discussion]{salim&narayanan20}. The log((\halpha+[NII])/[OIII]), $\log M_\star$, and 12+log(O/H) correlations are shown in Figure~\ref{fig:tau_property_correlations}. 

\begin{table}
\caption{Spearman's rank correlation coefficients, $\rho$, between $\tau_B^l$ and properties for ELGs in SDSS ($z<0.2$) \label{tab:Spearman_summary}} 
\begin{center}
\begin{tabular}{lcc}
 \hline \hline
\vspace{0.02in} $X$ & 
$\rho$ & $\tau_B^l$--$X$ polynomial fit \\ 
 \hline
log(\halpha) & 0.29 & ... \\
log(-EW(\halpha)) & 0.07 & ... \\
log(\halpha+[NII]) & 0.36 & ... \\
log([OIII]) & -0.33 & ... \\
log((\halpha+[NII])/[OIII]) & \textbf{0.75} & $0.0964+0.126X+0.116X^2$\\
log(\halpha/[OIII])  & \textbf{0.74} & $0.103+0.177X+0.122X^2$\\
$\log M_\star$ & \textbf{0.73} & $2.315-0.632X+0.0445X^2$\\
$\log M_{\star,tot}$ & \textbf{0.75} & $3.905-0.961X+0.0602X^2$\\
log($L$(\halpha))-$\log M_\star$ & -0.44 & ... \\
log($L$(\halpha+[NII]))-$\log M_\star$ & -0.38 & ... \\
log($L$([OIII]))-$\log M_\star$ & -0.68 & ... \\
log(SFR) & 0.68 & ... \\
log(sSFR) & 0.19 & ... \\
12+log(O/H) & \textbf{0.71} & $33.49-8.106X+0.4917X^2$\\ 
$\beta_\mathrm{phot}$ & 0.48$^\dagger$ & ... \\
\hline
 \end{tabular}
\end{center}
 \textbf{Notes.} Line fluxes/luminosities are observed values (i.e., not dust-corrected). Physical properties correspond to the \textit{fiber-only} region, however we also examine the relation with \textit{total} stellar mass $\log M_{\star,tot}$ for comparison to other work. Cases with $\rho>0.7$ are boldface. $^\dagger$This value is from \citet{battisti16}, which matched $\sim$10,000 SDSS galaxies with UV data from \textit{GALEX}.
\end{table}

\subsubsection{WISP+3D-HST ($z\sim1.3$)}
We explored binning the WISP+3D-HST sample by all of the properties listed in Table~\ref{tab:Spearman_summary} (adopting 4 bins; SFRs were based on SED fitting), as well as the observed luminosity at 2800\AA, $L(2800\text{\AA})_\mathrm{phot}$, the offset from the MS relation based on observed \halpha\ ($\log (\mathrm{SFR}(\mathrm{MS})/\mathrm{SFR}(\mathrm{H}\alpha_\mathrm{obs})$), and the offset from the MS relation based on observed UV luminosity ($\log (\mathrm{SFR}(\mathrm{MS})/\mathrm{SFR}(L(2800\text{\AA})_\mathrm{phot})$). We find that binning by the parameters mentioned above do not produce stacks where $\tau_B^l$ monotonically relates to the binned property (within uncertainty), with the exception of bins based on log((\halpha+[NII])/[OIII]), $\log M_\star$, and $\beta_\mathrm{phot}$. The line ratios for these stacks are listed in Table~\ref{tab:stack_analysis}. As a reminder, for this Section and the remainder of the paper, we only consider galaxies with $\log M_\star < 10.2$ ($N=911$).

\begin{figure}
\begin{center}
\includegraphics[width=0.45\textwidth]{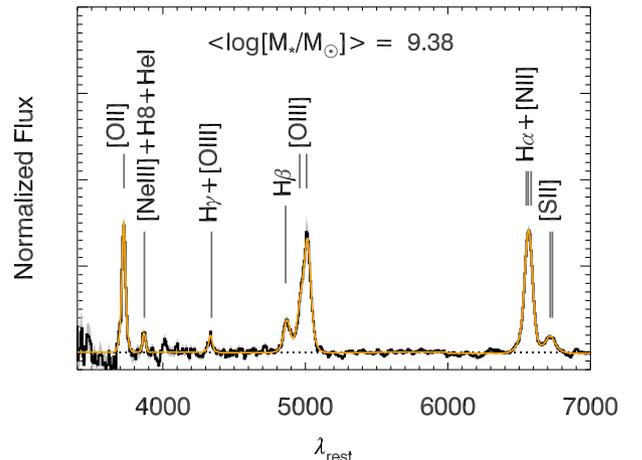}
\end{center}
\vspace*{-0.5cm}
\caption{Stacked spectrum normalised to \halpha+[NII] for a single bin of 65 galaxies at $1.35<z<1.5$ in WISP ($\log M_\star < 10.2$) with G102+G141, providing full coverage from [OII] to \halpha. The spectral fit (Section~\ref{method_stacking_fitting}) is shown as the orange line. The metallicity (based on R23) and Balmer decrement can be simultaneously measured for this stack (listed in Table~\ref{tab:stack_analysis}).  
 \label{fig:metal_stack}}
\end{figure}

Unfortunately, only $\sim$15\% of the sample (and only WISP sources) have simultaneous [OII] coverage for characterising metallicities via the R23\footnote{R23 = ([OII]$\lambda\lambda$3726, 3729 + [OIII]$\lambda\lambda$4959, 5007)/\hbeta} diagnostic. We construct a single bin of 65 galaxies at $1.35<z<1.5$ in WISP ($\log M_\star < 10.2$) with G102+G141, providing full coverage from [OII] to \halpha, where metallicities can simultaneously be determined with Balmer decrements. Following \citet{henry21}, we use the R23 diagnostic based on the calibration from \citet{curti17}. The stacked spectrum and emission line fits are shown in Figure~\ref{fig:metal_stack} and the line ratios are listed in Table~\ref{tab:stack_analysis}. 

We show the WISP+3D-HST stack values compared to SDSS in Figure~\ref{fig:tau_property_correlations}. The results from SDSS and  WISP+3D-HST indicate that correlations between $\tau_B^l$-$\log M_\star$ and $\tau_B^l$-log((\halpha+[NII])/[OIII]) are evident at both $z<0.2$ and $z\sim1.3$. Given that only one bin is available with metallicities in our current sample, we are unable to test if the $\tau_B^l$-metallicity correlation is present at both redshifts. We discuss the redshift evolution of these relations in Section~\ref{proxy_compare}.

\setlength{\tabcolsep}{2.5pt}
\begin{table*}
\caption{Summary of average properties and line ratios for WISP+3DHST final sample ($\log(M_\star/M_\odot) <10.2$) \label{tab:stack_analysis}} 
\begin{center}
\begin{tabular}{cccccccccccccc}
 \hline \hline
\vspace{0.02in} \multirow{2}{*}{$N$} & 
\multirow{2}{*}{$\left\langle X \right\rangle$} & 
\multirow{2}{*}{$\left\langle \log M_\star \right\rangle$} & 
\multicolumn{2}{c}{$\left\langle \log \mathrm{SFR} \right\rangle$} & 
\multirow{2}{*}{$\left\langle \beta_\mathrm{phot} \right\rangle$} & 
\multirow{2}{*}{$\dfrac{\mathrm{[OIII]}}{\mathrm{H}\beta}$}  & 
\multirow{2}{*}{$\dfrac{\mathrm{[OIII]}}{\mathrm{H}\beta_\mathrm{corr}}$} & 
\multirow{2}{*}{$\dfrac{(\mathrm{H}\alpha\mathrm{+[NII]})}{\mathrm{[OIII]}}$} & 
\multirow{2}{*}{$\dfrac{(\mathrm{H}\alpha\mathrm{+[NII]})}{\mathrm{H}\beta}$} & 
\multirow{2}{*}{$\dfrac{\mathrm{H}\alpha}{\mathrm{H}\beta}$} & 
\multirow{2}{*}{$\dfrac{\mathrm{H}\alpha_\mathrm{corr}}{\mathrm{H}\beta_\mathrm{corr}}$}& 
\multirow{2}{*}{$\tau_B^l$} & 
\multirow{2}{*}{$\dfrac{\mathrm{[SII]}}{(\mathrm{H}\alpha\mathrm{+[NII]})}$} \\ 
 & & & $\mathrm{H}\alpha_\mathrm{corr}$ & SED  \\
 \hline
\multicolumn{14}{c}{$X\equiv \log [ (\mathrm{H}\alpha+\mathrm{[NII]})/\mathrm{[OIII]} ]$} \\
     228 & -0.38 &  8.84 &  0.47 &  0.42 & -1.98 &  7.51 $\pm$ 0.44 &  7.37 $\pm$ 0.51 &  0.45 $\pm$ 0.01 &  3.38 $\pm$ 0.20
 &  3.14 $\pm$ 0.21 &  3.11 $\pm$ 0.23 &  0.09 $\pm$ 0.08 &  0.12 $\pm$ 0.01 \\
     228 & -0.18 &  9.20 &  0.73 &  0.55 & -1.77 &  6.05 $\pm$ 0.35 &  5.80 $\pm$ 0.53 &  0.71 $\pm$ 0.01 &  4.32 $\pm$ 0.25
 &  3.90 $\pm$ 0.28 &  3.79 $\pm$ 0.38 &  0.28 $\pm$ 0.10 &  0.16 $\pm$ 0.01 \\
     228 &  0.00 &  9.45 &  1.02 &  0.68 & -1.64 &  5.17 $\pm$ 0.42 &  4.85 $\pm$ 0.56 &  1.06 $\pm$ 0.02 &  5.48 $\pm$ 0.44
 &  4.80 $\pm$ 0.48 &  4.59 $\pm$ 0.56 &  0.47 $\pm$ 0.12 &  0.17 $\pm$ 0.01 \\
     227 &  0.44 &  9.68 &  1.28 &  0.94 & -1.26 &  2.55 $\pm$ 0.16 &  2.38 $\pm$ 0.27 &  2.55 $\pm$ 0.07 &  6.50 $\pm$ 0.39
 &  5.45 $\pm$ 0.56 &  5.19 $\pm$ 0.70 &  0.60 $\pm$ 0.13 &  0.21 $\pm$ 0.01 \\
 \hline
\multicolumn{14}{c}{$X\equiv \log M_\star$} \\
     228 &  8.60 &  ... &  0.58 &  0.27 & -2.02 &  7.20 $\pm$ 0.54 &  7.14 $\pm$ 0.55 &  0.53 $\pm$ 0.01 &  3.83 $\pm$ 0.29
 &  3.61 $\pm$ 0.27 &  3.60 $\pm$ 0.28 &  0.23 $\pm$ 0.08 &  0.14 $\pm$ 0.01 \\
     228 &  9.14 &  ... &  0.65 &  0.54 & -1.80 &  5.55 $\pm$ 0.34 &  5.36 $\pm$ 0.37 &  0.73 $\pm$ 0.01 &  4.06 $\pm$ 0.24
 &  3.68 $\pm$ 0.22 &  3.60 $\pm$ 0.24 &  0.23 $\pm$ 0.07 &  0.16 $\pm$ 0.01 \\
     228 &  9.50 &  ... &  0.92 &  0.75 & -1.56 &  5.11 $\pm$ 0.26 &  4.80 $\pm$ 0.34 &  1.00 $\pm$ 0.02 &  5.10 $\pm$ 0.25
 &  4.43 $\pm$ 0.24 &  4.24 $\pm$ 0.29 &  0.39 $\pm$ 0.07 &  0.18 $\pm$ 0.01 \\
     227 &  9.94 &  ... &  1.20 &  1.02 & -1.27 &  3.62 $\pm$ 0.26 &  3.23 $\pm$ 0.41 &  1.80 $\pm$ 0.05 &  6.52 $\pm$ 0.45
 &  5.13 $\pm$ 0.45 &  4.70 $\pm$ 0.59 &  0.50 $\pm$ 0.13 &  0.19 $\pm$ 0.01 \\
 \hline
\multicolumn{14}{c}{$X\equiv \beta_\mathrm{phot}$} \\     
     228 & -2.26 &  8.87 &  0.73 &  0.35 & -2.26 &  7.15 $\pm$ 0.61 &  7.01 $\pm$ 0.69 &  0.60 $\pm$ 0.01 &  4.30 $\pm$ 0.36
 &  3.99 $\pm$ 0.37 &  3.95 $\pm$ 0.41 &  0.32 $\pm$ 0.10 &  0.13 $\pm$ 0.01 \\
     228 & -1.89 &  9.23 &  0.76 &  0.50 & -1.89 &  5.80 $\pm$ 0.40 &  5.52 $\pm$ 0.56 &  0.79 $\pm$ 0.02 &  4.60 $\pm$ 0.31
 &  4.14 $\pm$ 0.35 &  4.01 $\pm$ 0.44 &  0.34 $\pm$ 0.11 &  0.16 $\pm$ 0.01 \\
     228 & -1.58 &  9.43 &  0.88 &  0.70 & -1.58 &  4.93 $\pm$ 0.32 &  4.65 $\pm$ 0.50 &  1.00 $\pm$ 0.02 &  4.93 $\pm$ 0.31
 &  4.33 $\pm$ 0.37 &  4.16 $\pm$ 0.49 &  0.37 $\pm$ 0.12 &  0.20 $\pm$ 0.01 \\
     227 & -0.91 &  9.64 &  1.14 &  1.03 & -0.91 &  3.54 $\pm$ 0.19 &  3.36 $\pm$ 0.35 &  1.58 $\pm$ 0.04 &  5.59 $\pm$ 0.28
 &  4.72 $\pm$ 0.47 &  4.56 $\pm$ 0.59 &  0.47 $\pm$ 0.13 &  0.19 $\pm$ 0.01 \\
\hline
\multicolumn{14}{c}{$X\equiv 12+\log(\mathrm{O/H})^\ddagger$}  \\     
      65 & 8.34 &  9.38 &  1.27 &  1.17 & -1.01 &  4.95 $\pm$ 0.38 &  4.87 $\pm$ 0.39 &  0.96 $\pm$ 0.03 &  4.77 $\pm$ 0.34
 &  4.22 $\pm$ 0.43 &  4.19 $\pm$ 0.44 &  0.38 $\pm$ 0.11 &  0.18 $\pm$ 0.01 \\
\hline
 \end{tabular}
\end{center}
 \textbf{Notes.} Columns similar to Table~\ref{tab:stack_mass}. $^\ddagger$Based on the R23 calibration of \citet{curti17}; only available for WISP galaxies at $1.35<z<1.5$.
\end{table*}


For deriving the average dust attenuation curve of our sample, we opt to use stacks binned according to log((\halpha+[NII])/[OIII]) because it shows the most uniform and maximal spread in $\tau_B^l$ and its bins also share similar average ages based on sSFRs (discussed in Section~\ref{beta_tau}). This quantity also has the benefit of being entirely empirical and does not require any assumptions on line blending corrections or dust attenuation, with the latter being required to derive galaxy properties (i.e., avoiding the issue of circularity).


\subsection{Relationship between UV slope and Balmer optical depth}\label{beta_tau}
The first step in our analysis is to examine the relationship between the reddening on the stellar continuum and the reddening on the ionised gas, as traced by the UV slope ($\beta$) and Balmer optical depth ($\tau_B^l$), respectively. When making this comparison it is important to account for intrinsic differences between the stellar populations among the sample. This is important because of the well known age/dust degeneracy in which the reddening of a young, dusty stellar population can be similar to that of an old, dust-free population \citep[e.g.,][]{witt92,gordon97}. For example, SFGs experiencing more dust attenuation are also, on average, more massive (e.g., Figure~\ref{fig:tau_property_correlations}) and have slightly older average stellar population ages \citep[e.g.,][]{battisti16} relative to SFGs experiencing less dust attenuation.

A common way to mitigate this issue is to compare samples of galaxies with roughly similar average stellar population ages based on an age-sensitive diagnostic, such as the 4000\AA\ break \citep[e.g.,][]{battisti16} or sSFR \citep[e.g.,][]{reddy15}. We adopt the sSFR(\halpha$_\mathrm{corr}$) of our stacks as a crude measure of their average stellar population age. For the 4 bins based on log((\halpha+[NII])/[OIII]), the values of log($\langle$sSFR(\halpha$_\mathrm{corr})\rangle$/yr$^{-1}$) are -8.40, -8.44, -8.44, -8.40, respectively, indicating that they share  similar average stellar population ages. A notably larger difference occurs when binning by stellar mass (-8.02, -8.54, -8.48, -8.74, respectively), in agreement with expectations that more massive galaxies, on average, have older stellar populations. 

\begin{figure*}
\begin{center}
$\begin{array}{ccc}
\includegraphics[scale=0.85]{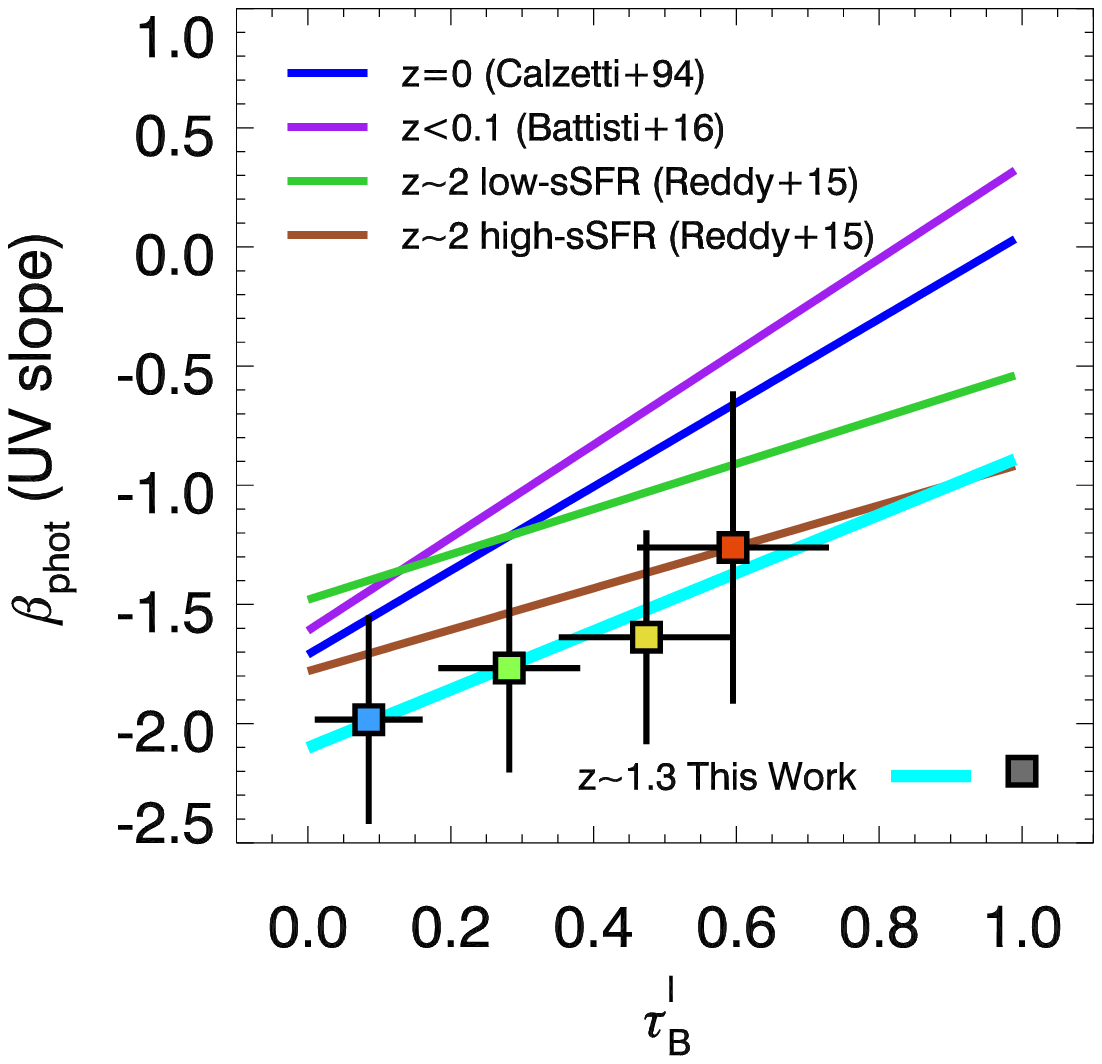} &
\includegraphics[scale=0.85]{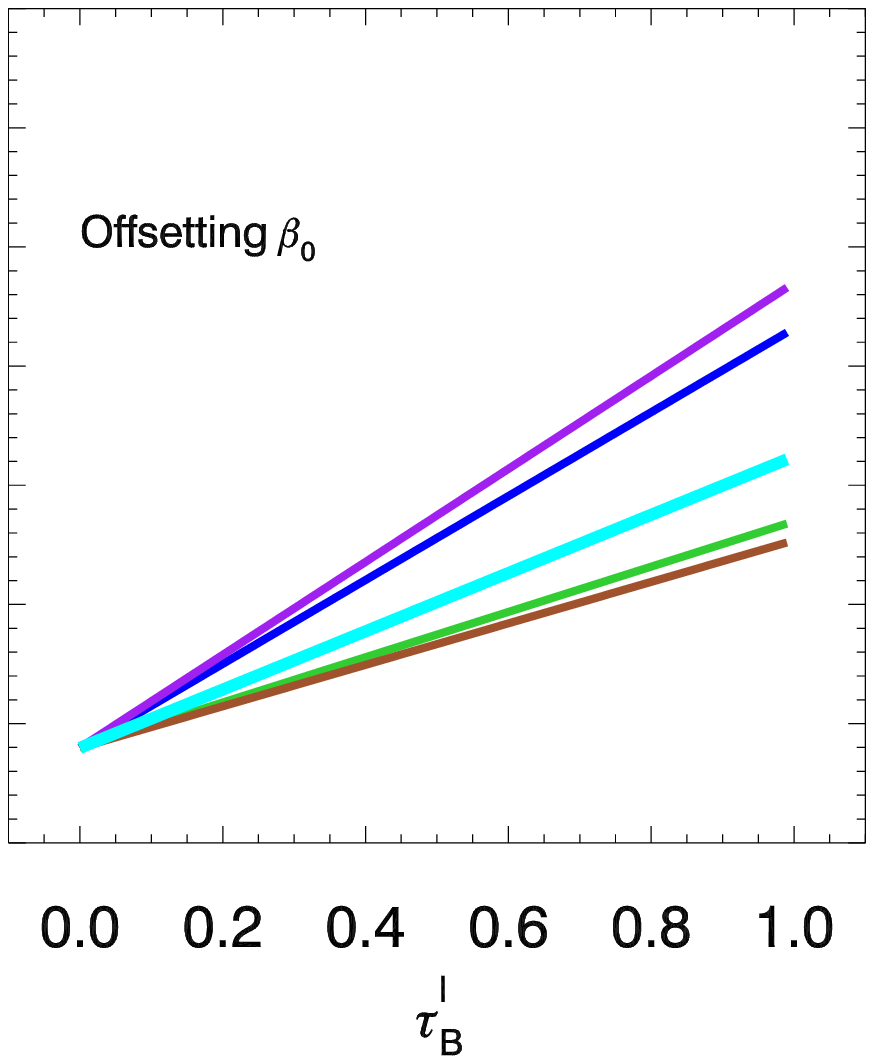} &
\end{array}$
\end{center}
\vspace{-0.3cm}
\caption{\textit{Left:} Relationship between the UV slope, $\beta_\mathrm{phot}$, and Balmer optical depth, $\tau_B^l$, for stacks binned according to log((\halpha+[NII])/[OIII]) (squares), together with relationships from the literature. The x-axis errorbars represent the measurement uncertainty from the stacked data, whereas the y-axis errorbars represent the standard deviation of $\beta_\mathrm{phot}$ values among individual galaxies within that bin. Vertical offsets likely reflect differences in the intrinsic UV slope due to stellar populations (Section~\ref{method_attenuation}), such as age \citep[e.g., sSFR bins of][]{reddy15} and metallicity. The slope of our relation lies in-between those found at $z\sim0$ and $z\sim2$, suggesting the relationship evolves with redshift. This is easier to see in the \textit{right} panel, where the relations are all normalised to the same $\beta_\mathrm{0}$ value. 
 \label{fig:beta_tau}}
\end{figure*}

The $\beta$--$\tau_B^l$ relation we find for our stacked sample is shown in Figure~\ref{fig:beta_tau} and compared to literature relations at lower and higher redshifts. For the \citet{reddy15} sample, the low-sSFR bin spans $-9.60\leq \log($sSFR/yr$^{-1}) \leq -8.84$ and high-sSFR bin spans $-8.84\leq \log($sSFR/yr$^{-1})\leq -8.00$. For reference, our sample has $\langle\log($sSFR/yr$^{-1})\rangle \sim -8.4$. Given the scatter/uncertainty of our measurements, we make the simplest assumption of a linear relationship between these quantities and obtain a fit of
\begin{equation}\label{eq:beta_tau}
\beta_{\mathrm{phot}} = (1.22\pm0.26) \tau_B^l -(2.10\pm0.09) \,,
\end{equation}
where the normalisation reflects the intrinsic UV slope at zero attenuation on the ionised gas ($\beta_0 = \beta(\tau_B^l=0)$). As mentioned earlier, the intrinsic value of $\beta_0$ in stellar populations varies depending on the age, metallicity, binarity of massive stars (Section~\ref{method_attenuation}). By normalising all of the curves to the same $\beta_0$ value (Figure~\ref{fig:beta_tau}, \textit{Right}), it is easier to see that the slope of the $z\sim1.3$ relation (1.22) lies in-between those found at $z\sim0$ ($\sim$1.8) and $z\sim2$ ($\sim$0.9) and supports the hypothesis that this relationship evolves with redshift.

\subsection{Average Dust Attenuation Curve}
To empirically determine dust attenuation curves, we follow the methodology used by previous studies based on Balmer decrements \citep[e.g.,][]{calzetti94, reddy15, battisti16, shivaei20a}. This requires comparing the average SEDs of galaxies with differing amounts of dust attenuation, based on $\tau_B^l$, while also accounting for intrinsic differences between the stellar populations among the sample. The latter,  if unaccounted for, would lead to spurious inferences on dust attenuation curves due to older stellar populations being intrinsically redder than younger stellar populations. As mentioned in the previous section, we find that our log((\halpha+[NII])/[OIII]) share similar average stellar population ages based on sSFR(\halpha$_\mathrm{corr}$) such that we do not expect significant differences in the intrinsic SED of our different bins. We explored the possibility of binning the sample according to multiple parameters (e.g., both mass and log((\halpha+[NII])/[OIII])) but found that the degree of scatter in SED shapes for smaller bin sizes are prohibitively large relative to differences in balmer decrement from the stacked spectra and so it is not explored further in this work. Studies in which individual Balmer decrements are available will be better suited to subdivide the data for this type of analysis.

Assuming a linear $\beta$--$\tau_B^l$ relationship (Figure~\ref{fig:beta_tau}), the optical depth is expected to behave according to \citep{calzetti94}
\begin{equation}
\tau_{n,r}(\lambda) = -\ln \frac{F_n(\lambda)}{F_r(\lambda)} \,,
\end{equation}
where $\tau_{n,r}$ corresponds to the dust optical depth of a reddened SED $n$, with flux density $F_n(\lambda)$, relative to a reference (less reddened) SED $r$, with flux density $F_r(\lambda)$. From this relation, it is possible to determine the \textit{selective} attenuation, $Q_{n,r}(\lambda)$, 
\begin{equation}\label{eq:Q_def}
Q_{n,r}(\lambda) = \frac{\tau_{n,r}(\lambda)}{\delta \tau_{Bn,r}^l} \,,
\end{equation}
where $\delta \tau_{Bn,r}^l=\tau_{Bn}^l-\tau_{Br}^l$ is the difference between the Balmer optical depth between $n$ and $r$. The normalisation for the selective attenuation is arbitrary. Following previous work \citep[e.g.,][]{calzetti94}, we choose $Q_{n,r}(5500\mathrm{\AA})=0$ as the zero-point.

The selective dust attenuation curve is related to the \textit{total-to-selective} dust attenuation curve, $k(\lambda)$, through the following relation
\begin{equation}\label{eq:k_def}
k(\lambda)= fQ(\lambda)+R_V \,,
\end{equation}
where $f$ is a constant required to make $k(B)-k(V)=1$,
\begin{equation}\label{eq:f_def}
f_{n,r}= \frac{1}{Q_{n,r}(B)-Q_{n,r}(V)} \,,
\end{equation}
and $R_V\equiv \frac{A(V)}{A(B)-A(V)}$ is the total-to-selective attenuation in the $V$ band, which is the vertical offset of the curve from zero at 5500~\AA\ (i.e., $R_V\equiv k(V)$).

The $f$ term accounts for differences in the reddening between the ionised gas and the stellar continuum, often referred to as differential reddening, and can also be expressed as 
\begin{equation}\label{eq:f_def_ebv_ratio}
f_{n,r} = \frac{k(H\beta)-k(H\alpha)}{\Delta E(B-V)_{\mathrm{star}}/\Delta E(B-V)_{\mathrm{gas}}} \,,
\end{equation}
where $k(H\beta)$ and $k(H\alpha)$ are the values for the intrinsic \textit{extinction} curve of the galaxy and \textit{not} from the attenuation curve and $\Delta E(B-V)=E(B-V)_n-E(B-V)_r$ is the difference in reddening in each component between the bins being compared. In the event that the reference bin is unreddened, this simplifies to 
\begin{equation}\label{eq:f_def_ebv_ratio2}
f_{n,0} = \frac{k(H\beta)-k(H\alpha)}{E(B-V)_{\mathrm{star}}/E(B-V)_{\mathrm{gas}}} \,.
\end{equation}
Eq~(\ref{eq:f_def_ebv_ratio}) implies that the nebular lines experience extinction instead of attenuation, which is commonly assumed and appears to be a reasonable assumption at least for optical nebular lines \citep[e.g.,][]{reddy20}. It is worth emphasizing that this assumption regarding the intrinsic extinction curve shape is only necessary if trying to quantify the differential reddening and that the derivation of the dust attenuation curve itself is entirely empirical and is independent of this assumption.

\begin{figure*}
\begin{center} $
\begin{array}{cc}
\includegraphics[scale=0.63]{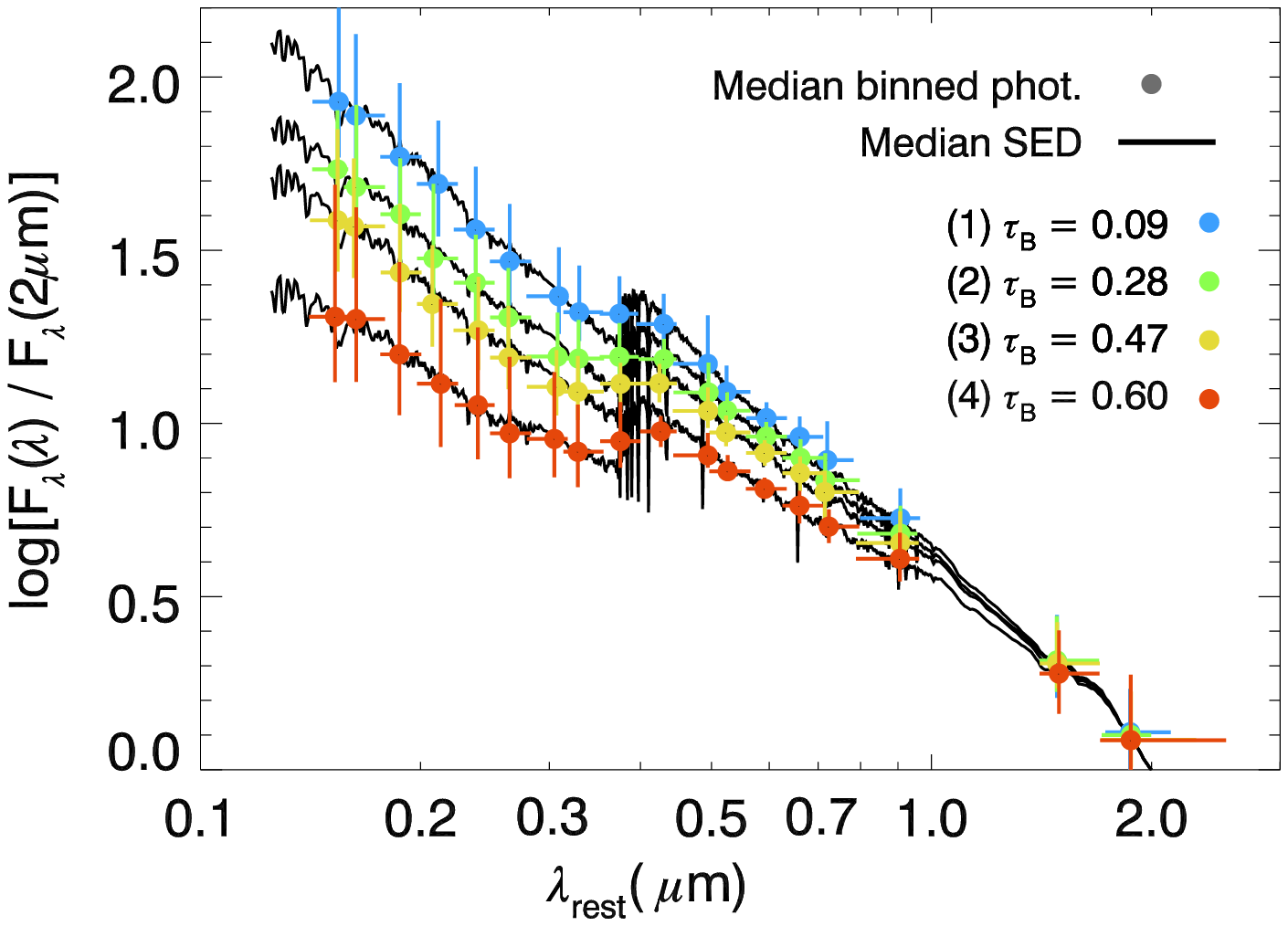} &
\includegraphics[scale=0.49]{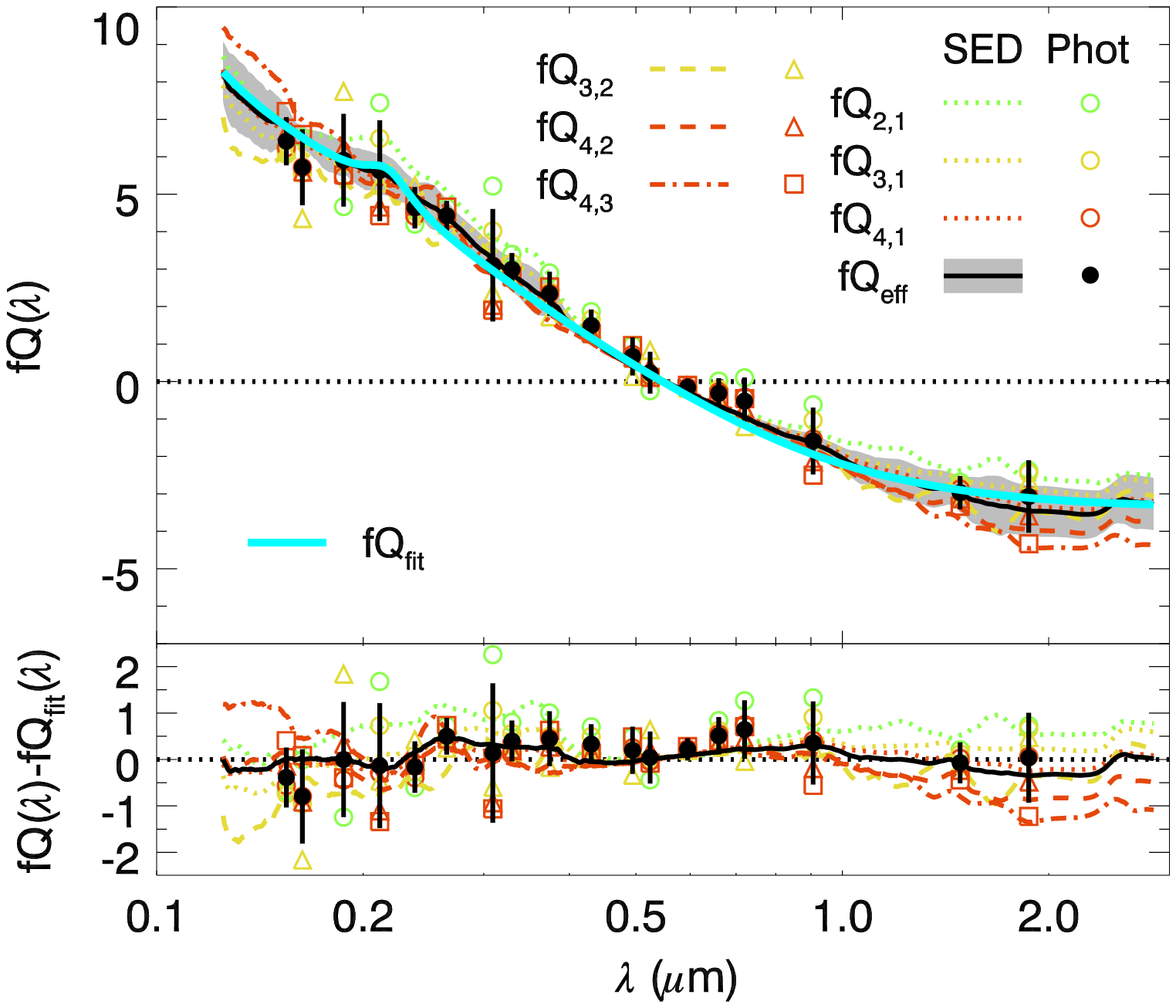} \\
\end{array}$
\end{center}
\vspace{-0.2cm}
\caption{\textit{Left:} Average SEDs for galaxies binned according to log((\halpha+[NII])/[OIII]). Symbols are the median values of all $S/N>3$ photometry divided in equal width (in log-units) bins of $\Delta\log(\lambda/\micron)=0.05$, with the exception of the IRAC region ($\lambda_\mathrm{rest}>1.4\micron$), which is divided into two bins. This provides $\sim$100--300 photometric points in each bin. The solid lines are the median best-fit SEDs from the \magphys\ fits.  A trend is evident of increasing reddening on the stellar continuum with increasing values of $\tau_B^l$ (ionised gas reddening; i.e., $\beta$--$\tau_B^l$ relation). The (1)--(4) values indicate the SED IDs. \textit{Right:} selective attenuation curve, $fQ_{n,r}(\lambda)$, based on comparing a given SED, $n$, to a reference SED, $r$, at lower $\tau_B^l$ (eq~\ref{eq:Q_def}). The solid black line and circles are the effective (average) curve, $fQ_{\mathrm{eff}}(\lambda)$ and the solid cyan line is our best fit. The gray shaded region denotes the 1$\sigma$ dispersion of $fQ_{n,r}(\lambda)$. The lower panel shows the residuals relative to the best fit. \label{fig:fQ_eff}}
\end{figure*}

We show the average SEDs for our binned sample and their corresponding Balmer decrements in Figure~\ref{fig:fQ_eff}, \textit{Left}. First, we determine the individual values for $f_{n,r}$ by fitting a third-order polynomial as a function of $x=1/\lambda$~($\micron^{-1}$) over the median photometric data in the range from 0.3--1.0$\mu$m. The resulting $f_{n,r}$ values for each SED pairing are summarised in Table~\ref{tab:fnr}, together with the differential reddening factor when assuming the ionised gas is reddened according to a MW extinction curve \citep{fitzpatrick19}. We find an average of $\langle f\rangle=3.80\pm1.12$, with the individual values showing a slight trend toward lower $f$ with increasing $n$, although $f_{3,2}$ is an outlier, which would correspond to increasing dust attenuation (also roughly to increasing stellar mass). A discussion of these differential reddening values in the context of other studies is presented in Section~\ref{diff_reddening_compare}.

\begin{table}
\caption{Summary of individual$f_{n,r}$ values  \label{tab:fnr}} 
\begin{center}
\begin{tabular}{ccc}
 \hline \hline
\vspace{0.02in} $n,r$ &
$f_{n,r}$ & 
$\Delta E(B-V)_{\mathrm{star}}/\Delta E(B-V)_{\mathrm{gas}}^\dagger$ \\ 
 \hline
2,1 & 3.78 & 0.31 \\
3,1 & 4.42 & 0.26 \\
4,1 & 3.42 & 0.34 \\
3,2 & 5.54 & 0.21 \\
4,2 & 3.41 & 0.34 \\
4,3 & 2.20 & 0.53 \\
 & $3.80\pm1.12$ & $0.31\pm0.11$ \\
 \hline
 \end{tabular}
\end{center}
 \textbf{Notes.} The last line shows the average and standard deviation among the sample. $^\dagger$This assumes the MW extinction curve as the intrinsic extinction curve (see eq~(\ref{eq:f_def_ebv_ratio})). 
\end{table}

The individual selective attenuation curves for the different pairings is shown in Figure~\ref{fig:fQ_eff}, \textit{Right}. It can be seen that each pairing has a similar shape, implying that adopting a single average selective attenuation curve is reasonable to characterise the entire population. We determine the effective attenuation curve, $fQ_{\mathrm{eff}}(\lambda)$, by taking the average value of $fQ_{n,r}(\lambda)$ found from considering both the median SEDs based on the \magphys\ best-fit SEDs and the photometry directly. These two methods provide consistent results within their error (scatter). We fit this average relation in two regimes. For the UV-optical regime ($0.125~\micron<\lambda\leq0.7~\micron$), we use a third-order polynomial as a function of $x=1/\lambda$~($\micron^{-1}$), together with a Lorentzian-like Drude profile to describe the 2175\AA\ feature \citep{fitzpatrick&massa90} 
\begin{equation}
D(\lambda) = \frac{E_b(x\,\gamma)^2}{(x^2-x_0^2)^2+(x\,\gamma)^2} \,,
\end{equation}
where $x_0=1/(0.2175\micron)$ and $\gamma=0.922\micron^{-1}$ are the MW values for the central wavelength and width of the feature, respectively \citep{fitzpatrick&massa07}, and $E_b$ is a free parameter for the bump intensity. For reference, the MW curve has an $E_b=3.30$ \citep{fitzpatrick99}, the average LMC and LMC2 supershell (30 Dor) curves have an $E_b=3.12$ and $1.64$, respectively \citep{gordon03}, and the SMC has a value of $E_b\lesssim0.39$ \citep{gordon03}. 

For the optical-NIR regime ($0.7~\micron<\lambda\leq2.85~\micron$), we use a separate third-order polynomial. The value of $R_V$ is determined by requiring that the total-to-selective attenuation approaches zero at long wavelengths, $k(\lambda\rightarrow\infty)\sim0$. A common assumption is to enforce this at $\lambda\sim2.85$~\micron\ \citep[e.g.,][]{calzetti97b, reddy15, battisti16}, which we also assume to avoid large extrapolations beyond the range of our photometry ($\lambda_\mathrm{rest}\lesssim2$~\micron). This provides a value of $R_V=3.26\pm0.70$, where the error corresponds to the 1$\sigma$ dispersion of $fQ_{n,r}(2.85\micron)$. 

\begin{figure*}
\begin{center} $
\begin{array}{cc}
\includegraphics[scale=0.58]{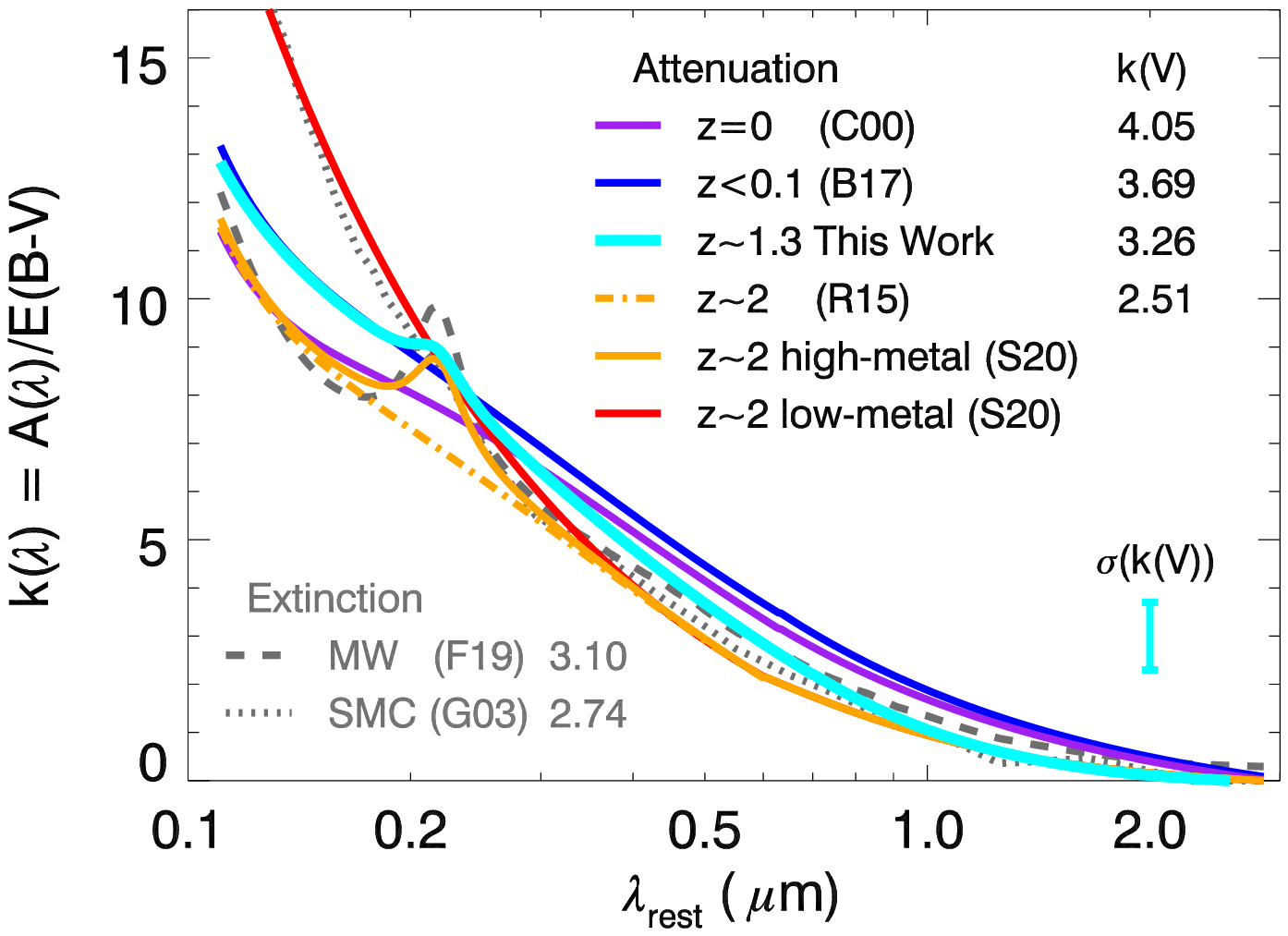} &
\includegraphics[scale=0.58]{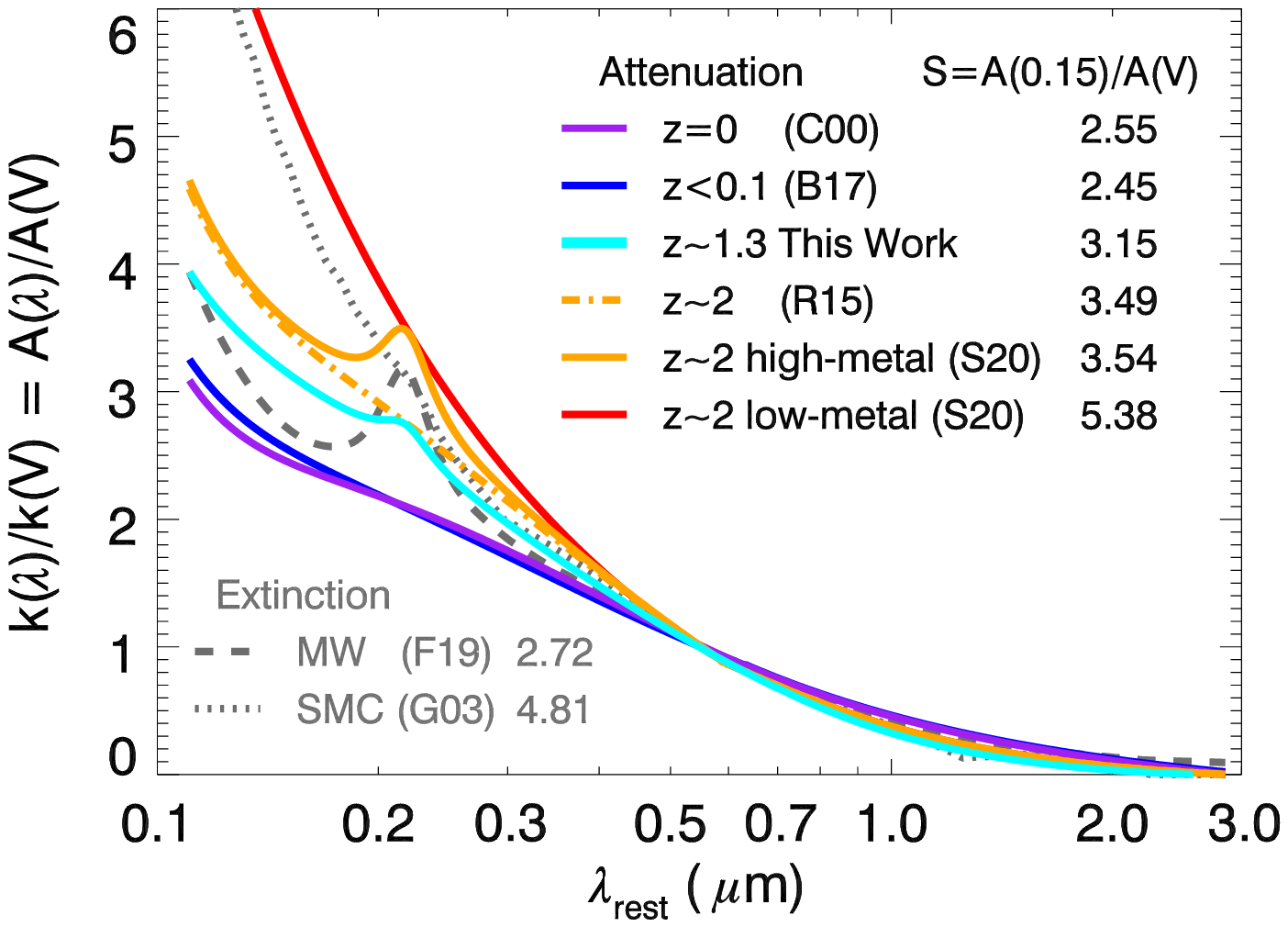} \\
\end{array}$
\end{center}
\vspace{-0.6cm}
\caption{\textit{Left:} Total-to-selective dust attenuation curve, $k(\lambda)$, for our $z\sim1.3$ sample (cyan line) compared to other studies using the same Balmer decrement technique at other redshifts (colour lines), with the curve normalisation, $k(V)$, for each also denoted. The extinction curves for the MW and SMC are also shown for reference (gray lines). Attenuation references are \citet{calzetti00, battisti17a, reddy15, shivaei20a} and extinction references are \citet{fitzpatrick19, gordon03}. Note, \citet{shivaei20a} adopt the \citet{reddy15} curve at $\lambda>0.6$~\micron, hence their $k(V)$ is nearly identical and not labelled. The value of $\sigma(k(V))$ for our curve is shown in the lower-right. \textit{Right:} Similar to left, but showing the normalised dust attenuation curve, $k(\lambda)/k(V)$, with the UV curve slope, $S=A(0.15)/A(V)$, for each also denoted. The values of $k(V)$ and $S$ at $z\sim1.3$ lie in-between those at $z\sim0$ and $z\sim2$, supporting the notion that the average values of these quantities evolve with redshift.  \label{fig:k_curves}}
\end{figure*}

The resulting average total-to-selective dust attenuation curve for our sample at $z\sim1.3$ is
\begin{equation}\label{eq:k_lambda}
  \begin{array}{ccr} 
  \multicolumn{3}{c}{k(\lambda)= 3.80\,(-1.542 + 1.046x - 0.123x^2 + 0.0063x^3)} \\
&  \multicolumn{2}{c}{+\dfrac{0.83 (x\,\gamma )^2}{((x^2-x_0^2)^2+(x\,\gamma )^2)}+3.26} \\ 
 & & 0.125~\micron\le\lambda<0.7~\micron \, \\ 
 & \multicolumn{2}{r}{= 3.80\,(-0.853 - 0.256x + 0.686x^2 - 0.157x^3)}\\ 
 & \hspace{30pt}  +3.26 & 0.7~\micron\le\lambda<2.85~\micron \,, 
\end{array}
\end{equation}
where $x=1/\lambda$~($\micron^{-1}$) and the 2175\AA\ feature has a strength of $E_b=0.83\pm0.08$. This corresponds to $\sim$25\% the strength of the feature in the average MW curve \citep[$E_b=3.3$;][]{fitzpatrick99}. This average attenuation curve is shown in two forms in Figure~\ref{fig:k_curves}. A detailed comparison of this curve in the context of the literature is presented in Section~\ref{curve_compare}.

\section{Discussion} 
\subsection{Redshift evolution of dust attenuation curves}\label{curve_compare}
A comparison between our $z\sim1.3$ average dust attenuation curve and other curves derived using a similar Balmer decrement method is shown in Figure~\ref{fig:k_curves}. In order of increasing redshift, these include \citet{calzetti00} based on a sample of $\sim$40 local starburst galaxies ($z=0$), \citet{battisti17a} based on a sample of $\sim$10,000 in SDSS, \citet{reddy15} based on $\sim$200 galaxies from the MOSDEF survey \citep{kriek15}, and \citet{shivaei20a} also based on the MOSDEF sample, but dividing it into two metallicity regimes ($8.2\lesssim 12 + \log(\mathrm{O/H}) <8.5$ and $12 + \log(\mathrm{O/H}) > 8.5$). Below we highlight a few notable trends that can be seen. We acknowledge that numerous additional attenuation curves have been derived based on other techniques, such pair-matching \citep[e.g.,][]{wild11}, SED-fitting \citep[e.g.,][]{conroy10, kriek&conroy13, scoville15, zeimann15}, and SED-fitting via energy-balance \citep[e.g.,][]{buat12, salim18, battisti20}. However, these can be challenging to directly compare between due to differing assumptions that must be made and we refer readers to the review of \citet{salim&narayanan20} for a detailed discussion. 

Three important quantities that describe dust curves are $k(V)$ (same as $R_{V}$), the UV curve slope $S=A(0.15)/A(V)$, and the 2175\AA\ bump strength $E_b$. In dust extinction curves, the values of these three quantities has been directly attributed to differences in the size distribution and chemical composition of dust grains \citep{weingartner&draine01, draine03}. In contrast, linking these quantities to intrinsic dust properties from dust attenuation curves is non-trivial owing to the additional geometric effects at play. This makes it significantly more difficult to assess whether differences in these quantities between galaxy samples are primarily driven by differences in the interstellar medium (ISM) or differences in geometry. However, it is worth noting that \textit{individual} galaxy dust attenuation curves may be subject to larger differences driven by geometric variation than for \textit{average} dust attenuation curves, due to the latter being derived from averaging over a large number of galaxies with varying star/dust geometries. Spatially-resolved studies of dust attenuation in nearby galaxies (e.g., using integral field spectrographs) may offer the best avenue to disentangle these effects. 

For our $z\sim1.3$ average dust attenuation curve, we find $k(V)=3.26$, $S=3.15$, and $E_b=0.83$. There is a clear trend of decreasing values of $R_{V}$ and increasing $S$ (steeper UV curve slopes), with increasing redshift. Of all galaxy properties, metallicity (through its link to dust grain composition) is among the most intuitive as a potential driver of the evolution in the \textit{average} dust attenuation curve shape due to its link to $S$ and $E_b$ in local dust extinction curves \citep{weingartner&draine01}, with the lower-metallicity SMC having a steeper curve and almost no 2175\AA\ feature ($S=4.81$ and $E_b\sim0$) compared to the higher-metallicty MW ($S=2.72$ and $E_b\sim3.1$). Indeed, the results of \citep{shivaei20a} separating the MOSDEF sample into low- and high-metallicity bins, via the [NII]/\halpha\ diagnostic \citep{pettini&pagel04}, supports this hypothesis. They find that their low-metallicity sample ($8.2\lesssim 12 + \log(\mathrm{O/H}) <8.5$) has a steeper curve with no significant 2175\AA\ feature relative to their high-metallicity sample ($12 + \log(\mathrm{O/H}) > 8.5$). Unfortunately, the low spectral resolution of the \hst\ grism data prevent the use of the [NII]/\halpha\ diagnostic and instead must rely on R23 \citep[e.g.,][]{henry21}, which is only available for a small subset of our  sample. 

No obvious trend in the strength of the 2175\AA\ feature ($E_b$) with redshift is evident, but we note that at low redshifts it has been difficult to establish the strength and variation of the 2175\AA\ feature because of the limited UV facilities capable of measuring it. For example, the most widely available low-$z$ data are from \galex, which has a wide NUV filter that diminishes the apparent strength of the feature by a factor of $\sim$2 \citep{battisti16}. Our average attenuation curve has $E_b=0.83$ or about $\sim$25\% of the strength found in the average MW curve. This is slightly lower than the strength of the feature found in the high-metallicity $z\sim2$ sample of \citet{shivaei20a}, which has a value of $E_b=1.53$ or about $\sim$50\% of the MW, but comparable to findings based on SED-fitting techniques for galaxies at low-$z$ \citep[e.g.,][]{salim18}, intermediate-$z$ \citep[e.g.,][]{buat12, battisti20}, and high-$z$ \citep{scoville15}. This is also comparable to \citet{noll09a}, who measured bump strengths ranging from $0.5\lesssim E_b\lesssim1.0$ by stacking rest-frame UV spectroscopy for galaxies at $1<z<2.5$. This suggests that the average strength of this feature may not evolve significantly with redshift. However, we note that some studies at $z\sim2$ have suggested that this feature is weak or not required in average dust attenuation curves \citep{reddy15, zeimann15, salmon16}. This disagreement may be due to differing selection effects on the galaxy samples used but it is also important to note that most constraints, with the exception of \citet{calzetti94} and \citet{noll09a}, are based on photometry with varying degrees of coverage in the 2175\AA\ region.  

We briefly highlight a couple additional factors that have been suggested to impact the 2175\AA\ feature. First, it is possible that strong UV radiation fields can process and/or destroy the dust grains that produce the 2175\AA\ feature and that this could play a more important role than metallicity \citep[e.g.,][]{clayton00, gordon03}. The strongest evidence for this claim is from \citet{clayton00}, who examined UV extinction (via spectroscopy) toward distant stars in the MW that reside in lower density regions, with environmental conditions more similar to the Magellanic Clouds, relative those used to derive the standard MW extinction curve and found a trend of decreasing bump strength with increasing steepness in the UV extinction curve (but still with MW metallicity). \citet{salim&narayanan20} explored this further and found that the bump strength across MW, LMC, and SMC sightlines show a positive correlation with optical extinction $A_V$ (a proxy of dust column density). The destruction of the 2175\AA\ carrier has been used to explain the absence of this feature from the starburst attenuation curve \citep{calzetti01} where the feature is conclusively ruled out based on UV spectroscopy from the $IUE$ but also shows a very shallow UV slope (counter to trends seen for extinction curves). Second, it is also expected that scattered light from an extended (unresolved) region by a foreground dust screen has the effect of reducing the overall optical depth, flattening the attenuation curve, and diminishing the strength of the 2175\AA\ feature \citep{natta&panagia84,calzetti94}. This geometric effect is the leading hypothesis for why the starburst attenuation curve is so shallow relative to extinction curves. 

Future IFS surveys at intermediate redshifts, where the 2175\AA\ feature is shifted to observer-frame optical, will be able to assess the strength of the feature in a spatially resolved manner as a function of various galaxy and ISM properties. This has the potential to provide real progress on the nature of the 2175\AA\ feature in galaxies. 

\subsection{Redshift evolution of differential reddening}\label{diff_reddening_compare}
In the local universe, it is found that ionised gas experiences roughly twice as much reddening as the stellar continuum, on average \citep[\EBVratioavg$\ \sim0.5$; e.g.,][]{calzetti00, kreckel13, battisti16}. This effect, referred to as differential reddening, is attributed to young stars (producing the ionised gas) in star forming regions being embedded in more dust than older stars in the diffuse ISM \citep[e.g.,][]{charlot&fall00}. At higher redshifts there is larger variation in the measured differential reddening \citep[e.g.,][see their Table 1 for compilation]{puglisi16, shivaei20a}, which may partly be driven by selection effects.  Studies have suggested that the differential reddening factor might approach unity as the sSFR of a galaxy increases because a larger fraction of the total stellar light will come from young stars that see dust in both the `birth cloud' and `diffuse ISM' components \citep[see Section~8.2.1 of][]{salim&narayanan20}.

Our results in Table~\ref{tab:fnr} may imply that larger differential reddening is occurring for less-dusty galaxies (less-massive) than more-dusty galaxies (more-massive), but this depends on the assumption for the same intrinsic extinction curve among the sample (at Balmer wavelengths). Using $\sim$500 galaxies at $1.4<z<2.6$ in the MOSDEF survey, \citet{reddy20} find that the nebular extinction curve in the Balmer region is consistent in shape to the MW extinction curve, suggesting this may be a reasonable assumption. As the average sSFR of our log((\halpha+[NII])/[OIII]) bins are roughly similar, this trend may suggest that larger differential reddening occurs at lower $\tau_B^l$ and moves closer to unity as $\tau_B^l$ increases. This is consistent with \citet{shivaei20a}, which found larger differential reddening in their low-metallicity bin (typically lower mass; \EBVratioavg=0.38) than their high-metallicity bin (typically higher mass; \EBVratioavg=0.72), although noting that they do not find significant differences when separating the sample by stellar mass. Recently, \citet{rodriguez-munoz21} also found a trend of larger differential reddening on \halpha\ for less-dusty galaxies than more-dusty galaxies, based on $A_\mathrm{UV}$, in a sample of $\sim$700 [OII] emitters at $0.3\lesssim z\lesssim 1.5$ (noting their definition of $f$ differs from ours; they use $f=E(B-V)_{\mathrm{star}}/ E(B-V)_{\mathrm{gas}}$). Deeper surveys extending to lower mass regimes, such as those planned with \jwst, and/or spatially resolved studies in the local universe will provide new insight into the nature of differential reddening.

\subsection{Redshift evolution of dust attenuation proxies}\label{proxy_compare}
Obtaining direct estimates of dust attenuation in galaxies are observationally expensive, requiring either spectroscopy for dust-sensitive line diagnostics (e.g., Balmer decrement) or rest-frame IR measurements. Both of these are extremely difficult to obtain at high redshifts. Currently, ALMA is the only facility capable of providing IR measurements at high redshifts, and even then only for small samples of galaxies. As a result, it is advantageous to determine what properties are closely linked to attenuation and can serve as a suitable proxy when no spectroscopy or IR data are available. 

The most commonly adopted proxy for dust attenuation is the UV slope, $\beta$, due to the ease with which it can be measured at high redshifts. For starburst galaxies, a tight correlation exists between the infrared excess, $IRX=L_{\rm{IR}}/L_{\rm{UV}}$, a proxy for the total dust attenuation, and $\beta$ \citep{meurer99}. However, the $IRX-\beta$ relation has been shown to break down as one moves from starburst galaxies to more ``normal'' SFGs \citep[e.g.,][]{kong04, buat05, hao11, capak15, pope17, salim&boquien19}. This break down is attributed to degeneracies in this relation due to variations in intrinsic dust extinction curves, metallicity, stellar population age, and/or the star-dust geometry \citep[e.g.,][]{salim&boquien19, salim&narayanan20, shivaei20b}. Unfortunately, for the current sample we do not have the necessary data to examine the nature of the $IRX-\beta$ relation at $z\sim1.3$. 

Several studies have also suggested that there may be a universal relationship between Balmer decrements (or Balmer optical depths) and galaxy stellar masses \citep[i.e., $\tau_B^l$-$\log M_\star$ relation;][]{dominguez13, kashino13, shivaei20a, villa-velez21, shapley21}. Most literature values appear to follow the \cite{garn&best10} relation found for SDSS galaxies, which is based on \textit{total} stellar masses ($\log M_{\star,\mathrm{tot}}$). However, as shown in Figure~\ref{fig:tau_property_correlations}, there is a constant offset between these relationships whether one considers the fiber-region stellar mass or the total mass ($\log M_{\star,\mathrm{tot}}\sim \log M_\star+0.5$). We believe that since the SDSS Balmer decrement measurements are based on the fiber-region, it is more intuitive to compare to the stellar mass within the same fiber-region because comparing to total mass would implicitly assume there is no radial dependence on Balmer decrements. Radial dependencies on Balmer decrements have been found at both $z\sim0$ \citep[e.g.,][]{greener20} and $z\sim1.4$ \citep[e.g.,][]{nelson16a}, which suggests that care should be taken to ensure consistency between the measurements being compared. However, this seems to contrast with the fact that the correlation strength between $\tau_B^l$-$\log M_{\star,\mathrm{tot}}$ is not lower than that found for the fiber-region stellar mass (see Table~\ref{tab:Spearman_summary}).
This highlights that future studies of galaxies that span a wider redshift range are needed to establish the extent to which the stellar mass-Balmer decrement relation is universal. This is the subject of an ongoing WISP project (Alavi et al. in prep.).

There also appears to be a significant correlation between $\tau_B^l$ and log((\halpha+[NII])/[OIII]) for both SDSS and our $z\sim1.3$ sample, but which appear offset and to differ in functional form. We test if these changes are most likely driven by an evolution in ionised gas properties (e.g., metallicity, ionisation parameter). To model this, we evolved the line ratios of SDSS galaxies, keeping their masses and $\tau_B^l$ values fixed, to $z=1.33$ using the redshift- and mass-dependent evolution of the [NII]/\halpha\ ratio from \citet{faisst18}. We then combined this with the redshift-dependent evolution of the BPT locus ([OIII]/\hbeta\ vs. [NII]/\halpha) based on the relations from \citet{faisst18} and \citet{kewley13a}. The expected changes from the $z\sim0$ SDSS $\tau_B^l$-log((\halpha+[NII])/[OIII]) relation are shown in Figure~\ref{fig:tau_property_correlations} and show remarkable similarity to the observed trend from our sample.

Finally, we examine the $\tau_B^l$-$12+\log(\mathrm{O/H})$ relation. Assuming a non-evolving $\tau_B^l$-$\log M_\star$ relation, an offset in $\tau_B^l$-$12+\log(\mathrm{O/H})$ would be expected due to the evolving mass-metallicity relation. In Figure~\ref{fig:tau_property_correlations}, we can see that our single metallicity bin (based on R23; $z\sim1.4$) does appear offset from the $z\sim0$ SDSS R23 metallicity relation, which tends toward lower metallicities relative to those inferred from \citet{tremonti04}. We refer readers to \citet{kewley&ellison08} for more details on systematic differences between metallicity diagnostics. \citet{henry21} find that the R23 mass-metallicity relation at $1.3<z<2.3$ is offset lower from the $z\sim0$ R23 relation by $-0.3$~dex in $12+\log(\mathrm{O/H})$ for a fixed stellar mass ($\tau_B^l$) and is demonstrated in Figure~\ref{fig:tau_property_correlations}. We note that the average mass and metallicity of our stack ($\log M_\star=9.38$, $12+\log(\mathrm{O/H})=8.34$) is consistent with the \citet{henry21} R23 mass-metallicity relation. The position of our stack is in qualitative agreement with the expected shift direction, but it appears to have a larger offset in $12+\log(\mathrm{O/H})$ ($\sim-0.45$~dex). Future surveys in which both metallicities and Balmer decrements can be simultaneously measured for a significant sample of galaxies are needed to establish whether a relation between $\tau_B^l$ and metallicity is present at higher redshifts.

\section{Conclusion}\label{conclusion}
Using $\sim$900 galaxies with $8\lesssim\log (M_\star /M_\odot)<10.2$ at $0.75<z<1.5$ in the \hst\ WFC3 IR Spectroscopic Parallel (WISP) and 3D-HST grism surveys, we examine the relationship between dust attenuation, as traced by the Balmer decrement, and various galaxy properties at $z\sim1.3$ (Figure~\ref{fig:tau_property_correlations}). We establish that the (\halpha +[NII])/[OIII] line ratio and stellar mass, which show a significant correlation with the Balmer optical depth ($\tau_B^l$) in SDSS, also appear to be good proxies for $\tau_B^l$ at $z\sim1.3$. The relation between $\tau_B^l$ and stellar mass appears roughly similar for both samples, indicative of minor redshift evolution, but the relation between $\tau_B^l$ and (\halpha +[NII])/[OIII] shows considerable difference. We demonstrate that this difference matches the evolution in these emission lines with redshift and can be attributed to the changing ISM physical conditions of galaxies at higher redshifts. The $\tau_B^l$-log((\halpha +[NII])/[OIII]) relation offers a useful means to roughly separate samples of galaxies by relative attenuation from grism data where only upper limits on \hbeta\ are available and/or \halpha\ is blended with [NII] such that Balmer decrements cannot be individually measured to quantify dust attenuation. 

Using stacked spectra, binned by (\halpha+[NII)]/[OIII], we derive the $\beta$-$\tau_B^l$ relation (eq~\ref{eq:beta_tau}; Figure~\ref{fig:beta_tau}) and average dust attenuation curve at $z\sim1.3$ (eq~\ref{eq:k_lambda}; Figure~\ref{fig:k_curves}). The $\beta$-$\tau_B^l$ relation at $z\sim1.3$ has a slope in-between previous relations at $z=0$ and $z\sim2$. Similarly, the values for the dust attenuation curve slope ($S=3.15$) and normalisation ($R_V=3.26$) lie in-between the values found for dust curves at $z=0$ \citep[$S=2.55$ and $R_V=4.05$;][]{calzetti00} and $z\sim2$ \citep[$S=3.49$ and $R_V=2.51$;][]{reddy15}. In contrast, the 2175\AA\ feature strength is comparable to values found at other redshifts from SED-fitting methods ($E_b=0.83$; $\sim$25\% that of the MW extinction curve), but which may have a strong metallicity dependence \citep[e.g.,][]{shivaei20a}. These results highlight that the average shape and normalisation of dust attenuation curve appear to evolve with redshift, but that the average strength of the 2175\AA\ feature may experience less evolution. 

We conclude by noting that larger statistical datasets of galaxies with direct Balmer decrements (not reliant on stacking to measure \hbeta) from upcoming grism surveys with \jwst, \euclid, and \rst\ offer the potential to examine changes in $R_{V}$, $S$, and $E_b$ as a function of galaxy properties in a more systematic manner. These datasets will also provide further guidance on the best proxies for dust attenuation when spectroscopic and/or IR diagnostics are unavailable.

\section*{Acknowledgements}
The authors thank the anonymous referee, whose suggestions helped to clarify and improve the content of this work.
Parts of this research were supported by the Australian Research Council Centre of Excellence for All Sky Astrophysics in 3 Dimensions (ASTRO 3D), through project number CE170100013. 
YSD acknowledges the support from National Key R\&D Program of China via grant No. 2017YFA0402704, the NSFC grants 11933003, and the China Manned Space Project with No. CMS-CSST-2021-A05. 
AJB thanks K. Grasha for comments that improved the paper. 
AJB thanks A. Faisst for helpful responses to inquiries pertaining to his paper.
AJB is also thankful for attending ASTRO 3D writing retreats that provided a helpful environment to complete portions of this manuscript. 
We acknowledge the invaluable labor of the maintenance and clerical staff at our institutions, whose contributions make our scientific discoveries a reality. 
This research was conducted on Ngunnawal Indigenous land.






\section*{Data Availability}
All of the photometric and spectroscopic data used in this paper are publicly available through data releases from the WISP, 3D-HST, and SDSS survey teams as described in Section~\ref{data_sample}. Other data products can be made available upon reasonable request to the first author.

\bibliographystyle{mnras}
\bibliography{AJB_bib}


\appendix
\section{Comparison to 3D-HST catalog}\label{3dhst_compare}

\begin{figure*}
\begin{center} $
\begin{array}{cc}
\includegraphics[scale=0.65]{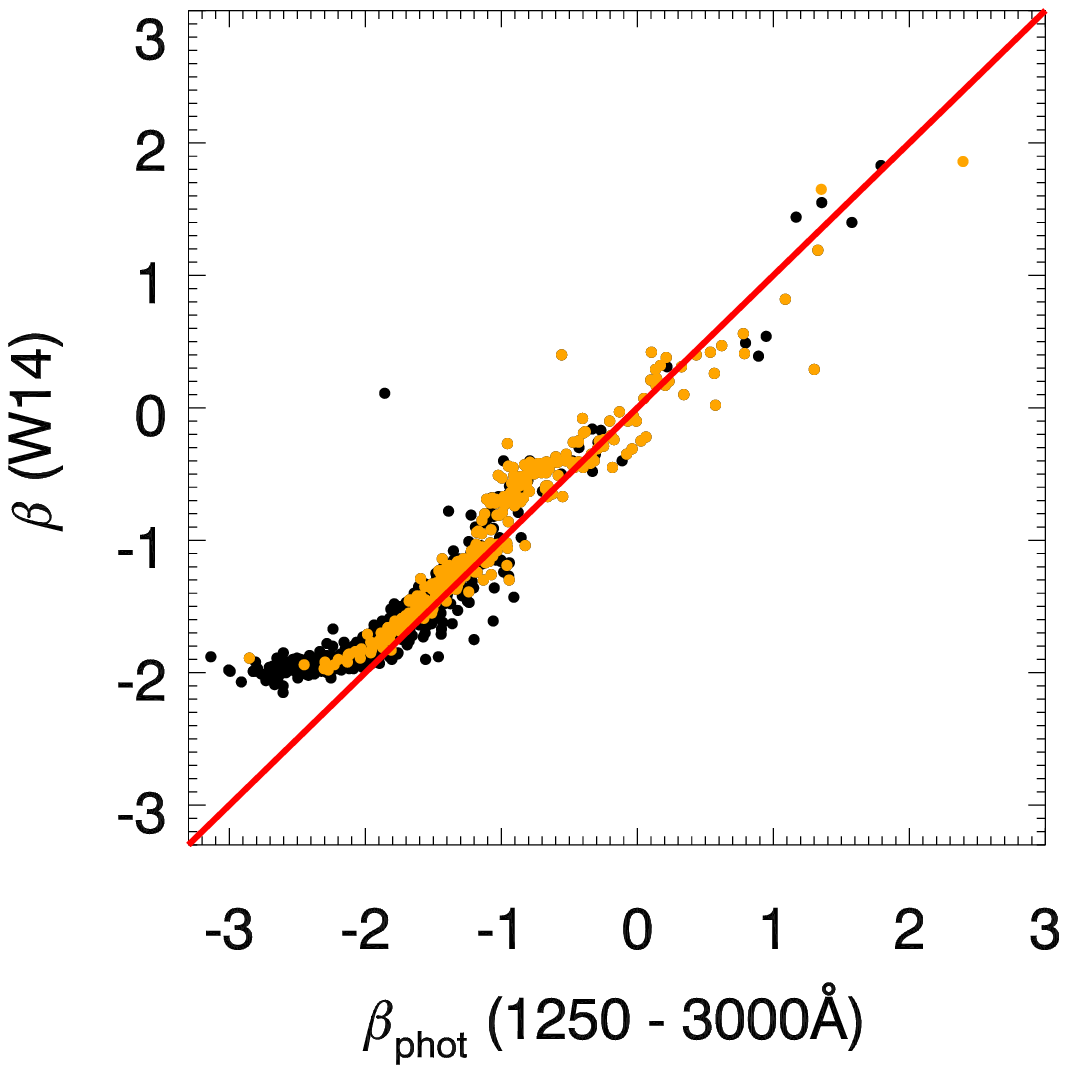} &
\includegraphics[scale=0.65]{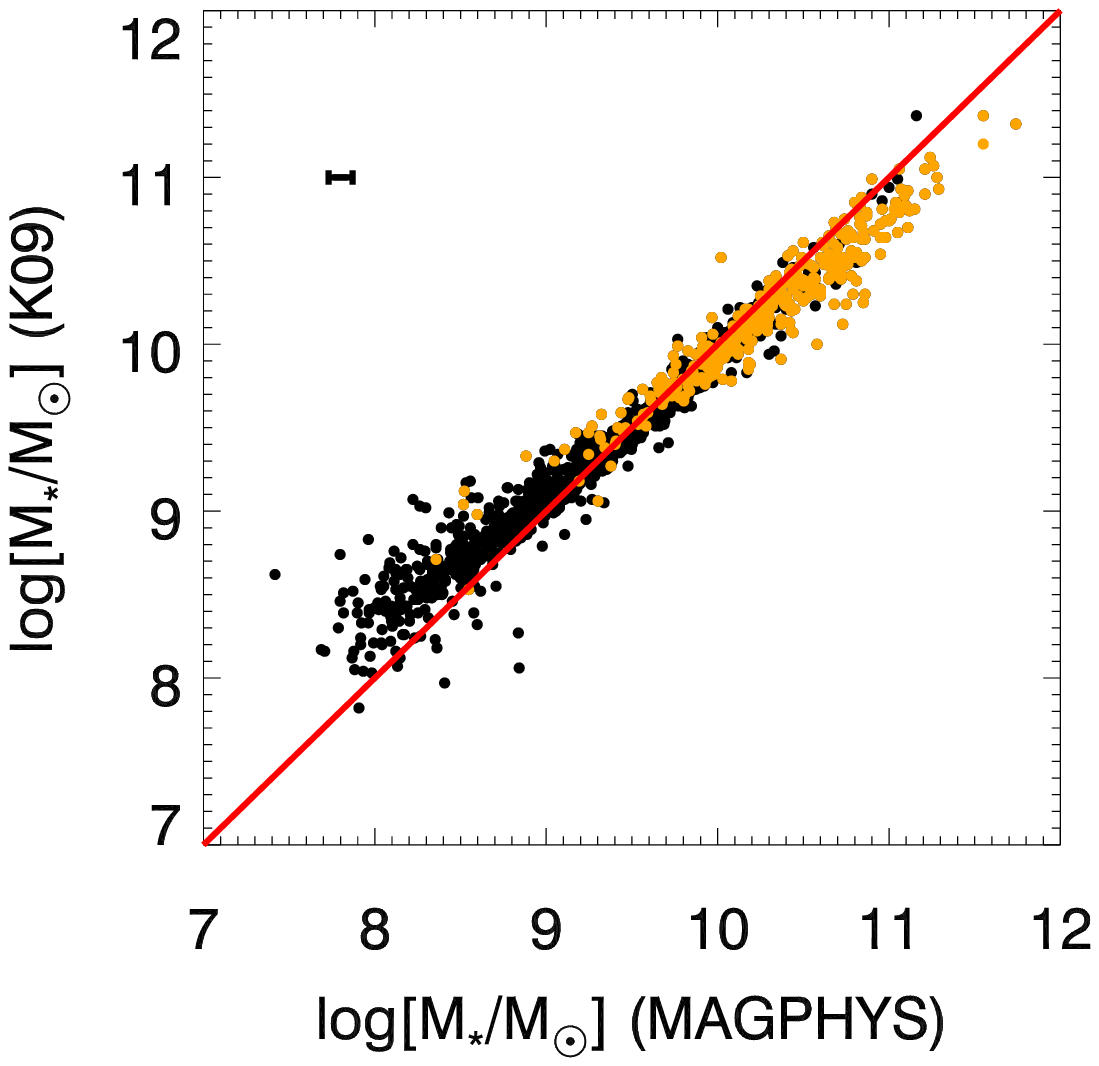} \\
\includegraphics[scale=0.65]{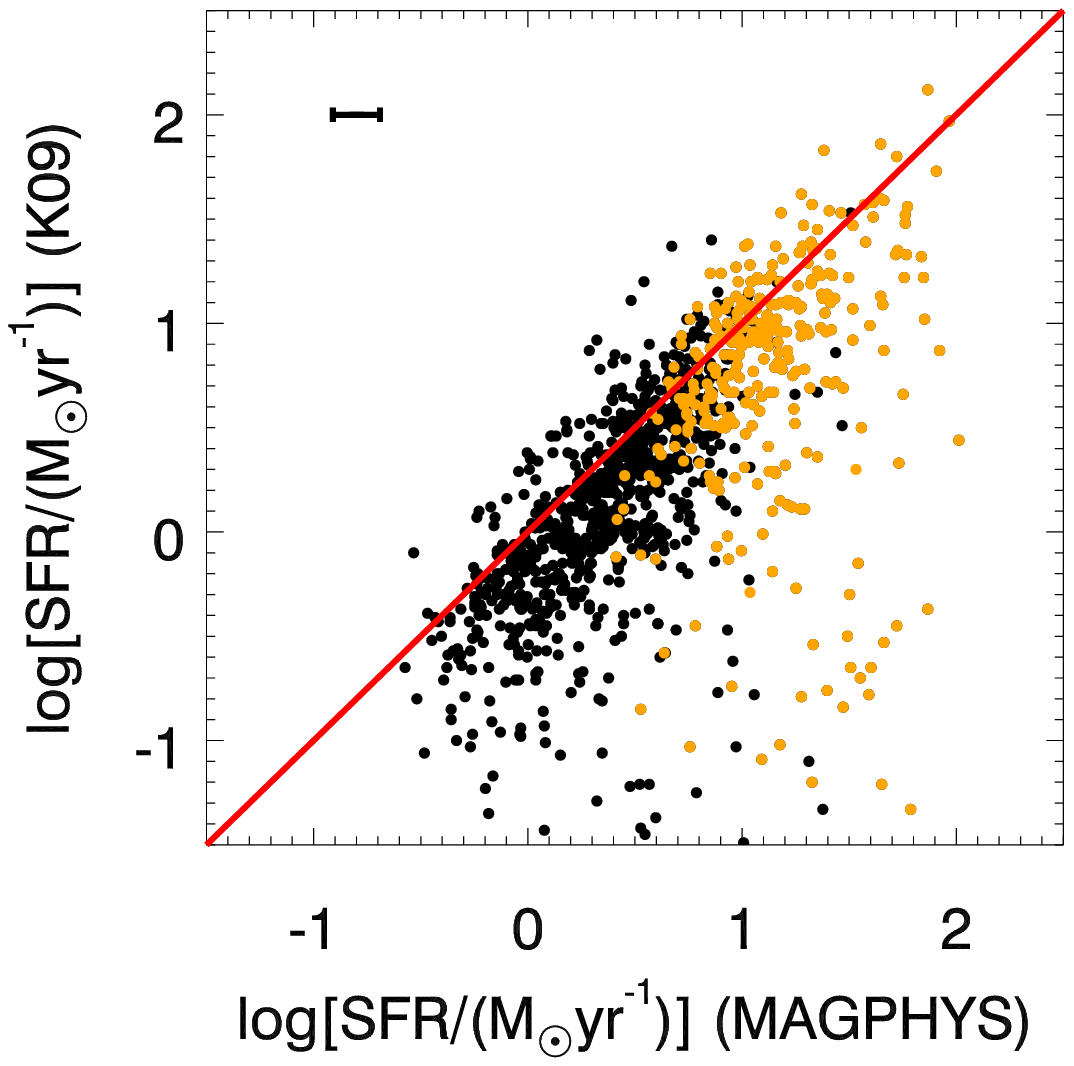} &
\includegraphics[scale=0.65]{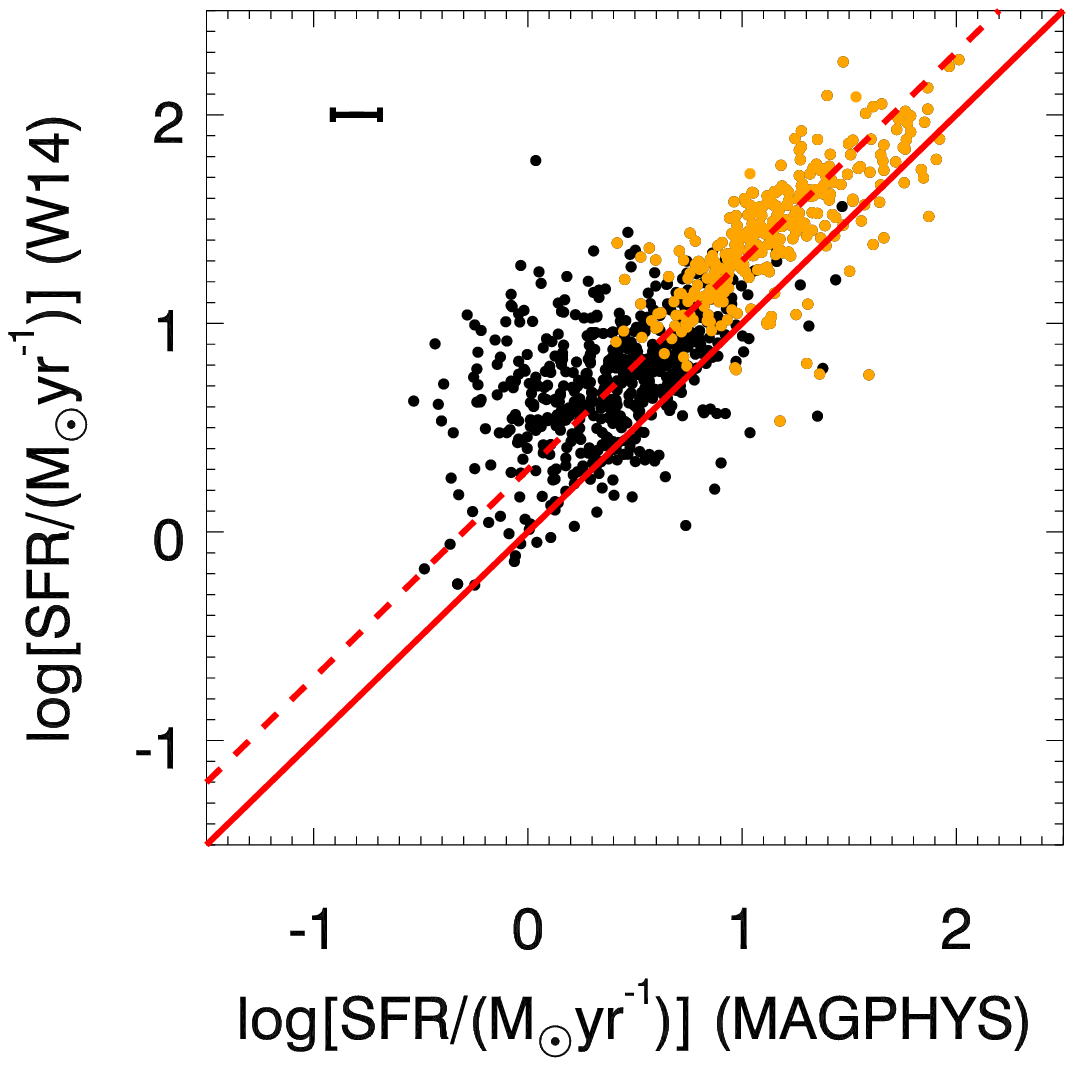} \\
\end{array}$
\end{center}
\caption{Comparison between our derived properties and those from the 3D-HST catalog \citep{skelton14}. Solid red lines show the 1:1 relations. Orange symbols denote 3D-HST galaxies with at least one IR band with $S/N$>3 (\spitzer\ 24\micron\ and/or Herschel; 26\% of sample). For \magphys\ quantites, the median 1$\sigma$ uncertainties are indicated in the upper-left of the panel (3D-HST catalog does not list errors).
\textit{Top-Left:} Our photometric-derived UV-slope, $\beta_\mathrm{phot}$, and those from \citet{whitaker14}. A minimum slope of $\beta\sim-2.15$ appears to occur for the 3D-HST catalog. 
\textit{Top-Right:} Our \magphys -derived stellar masses and those from FAST \citep[using only UV-near-IR photometry][]{kriek09}. 
\textit{Bottom-Left:} Our \magphys -derived SFRs and those from FAST \citep{kriek09}. 
\textit{Bottom-Right:} Comparison between \magphys -derived SFRs and those from \citet{whitaker14} based on UV+IR. The values of SFR(UV+IR) are systematically larger by $\sim$0.3~dex (dashed red line).
 \label{fig:3dhst_compare}}
\end{figure*}

We compare our measurements for UV slopes, stellar masses, and SFRs with those from the the 3D-HST catalog \citep{skelton14} in Figure~\ref{fig:3dhst_compare}. Most quantities show good agreement, with small offsets in some cases that we attribute to different assumptions. For the UV slopes, a slight feature is apparent at $\beta_\mathrm{phot}\sim-1$, but we find no obvious reason for this from visually inspecting our fits. For stellar masses, the deviation from 1:1 may be linked to differences in the dust attenuation prescription between the codes (\magphys\ uses \citet{charlot&fall00}; FAST used \citet{calzetti00}) and/or the SFH treatment. For SFRs, the large scatter in the derived values between the various methods highlights the difficulty in determining this quantity from only UV-NIR photometry. This informed our decision to not use SFRs based on the SED fitting as a parameter for binning or restricting our sample for our main analysis.

\bsp	
\label{lastpage}
\end{document}